\documentclass[twoside,twocolumn,9pt]{article}
\usepackage{extsizes}
\usepackage[super,sort&compress,comma]{natbib} 
\usepackage[version=3]{mhchem}
\usepackage[left=1.5cm, right=1.5cm, top=1.785cm, bottom=2.0cm]{geometry}
\usepackage{balance}
\usepackage{mathptmx}
\usepackage{sectsty}
\usepackage{graphicx} 
\usepackage{lastpage}
\usepackage[format=plain,justification=justified,singlelinecheck=false,font={stretch=1.125,small,sf},labelfont=bf,labelsep=space]{caption}
\usepackage{float}
\usepackage{fancyhdr}
\usepackage{fnpos}
\usepackage[english]{babel}
\addto{\captionsenglish}{%
 
}
\usepackage{amsmath}
\usepackage{wrapfig}
\usepackage{amssymb}

\usepackage{array}
\usepackage{droidsans}
\usepackage{charter}
\usepackage[T1]{fontenc}
\usepackage[usenames,dvipsnames]{xcolor}
\usepackage{setspace}
\usepackage[compact]{titlesec}
\usepackage{hyperref}
\usepackage[normalem]{ulem}

\usepackage{epstopdf}

\definecolor{cream}{RGB}{222,217,201}

\begin{document}

\pagestyle{fancy}
\thispagestyle{plain}
\fancypagestyle{plain}{
\renewcommand{\headrulewidth}{0pt}
}

\makeFNbottom
\makeatletter
\renewcommand\LARGE{\@setfontsize\LARGE{15pt}{17}}
\renewcommand\Large{\@setfontsize\Large{12pt}{14}}
\renewcommand\large{\@setfontsize\large{10pt}{12}}
\renewcommand\footnotesize{\@setfontsize\footnotesize{7pt}{10}}
\makeatother

\renewcommand{\thefootnote}{\fnsymbol{footnote}}
\renewcommand\footnoterule{\vspace*{1pt}%
\color{cream}\hrule width 3.5in height 0.4pt \color{black}\vspace*{5pt}} 
\setcounter{secnumdepth}{5}

\makeatletter 
\renewcommand\@biblabel[1]{#1} 
\renewcommand\@makefntext[1]%
{\noindent\makebox[0pt][r]{\@thefnmark\,}#1}
\makeatother 
\renewcommand{\figurename}{\small{Fig.}~}
\sectionfont{\sffamily\Large}
\subsectionfont{\normalsize}
\subsubsectionfont{\bf}
\setstretch{1.125} 
\setlength{\skip\footins}{0.8cm}
\setlength{\footnotesep}{0.25cm}
\setlength{\jot}{10pt}
\titlespacing*{\section}{0pt}{4pt}{4pt}
\titlespacing*{\subsection}{0pt}{15pt}{1pt}

\fancyfoot{}
\fancyfoot[LO,RE]{\vspace{-7.1pt}\includegraphics[height=9pt]{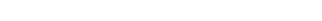}}
\fancyfoot[CO]{\vspace{-7.1pt}\hspace{13.2cm}\includegraphics{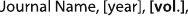}}
\fancyfoot[CE]{\vspace{-7.2pt}\hspace{-14.2cm}\includegraphics{head_foot/RF}}
\fancyfoot[RO]{\footnotesize{\sffamily{1--\pageref{LastPage} ~\textbar \hspace{2pt}\thepage}}}
\fancyfoot[LE]{\footnotesize{\sffamily{\thepage~\textbar\hspace{3.45cm} 1--\pageref{LastPage}}}}
\fancyhead{}
\renewcommand{\headrulewidth}{0pt} 
\renewcommand{\footrulewidth}{0pt}
\setlength{\arrayrulewidth}{1pt}
\setlength{\columnsep}{6.5mm}
\setlength\bibsep{1pt}

\makeatletter 
\newlength{\figrulesep} 
\setlength{\figrulesep}{0.5\textfloatsep} 

\newcommand{\topfigrule}{\vspace*{-1pt}%
\noindent{\color{cream}\rule[-\figrulesep]{\columnwidth}{1.5pt}} }

\newcommand{\botfigrule}{\vspace*{-2pt}%
\noindent{\color{cream}\rule[\figrulesep]{\columnwidth}{1.5pt}} }

\newcommand{\dblfigrule}{\vspace*{-1pt}%
\noindent{\color{cream}\rule[-\figrulesep]{\textwidth}{1.5pt}} }

\makeatother

\twocolumn[
 \begin{@twocolumnfalse}
{\includegraphics[height=30pt]{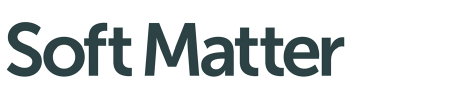}\hfill\raisebox{0pt}[0pt][0pt]{\includegraphics[height=55pt]{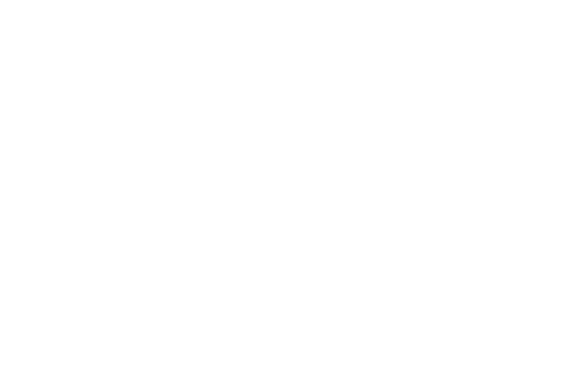}}\\[1ex]
\includegraphics[width=18.5cm]{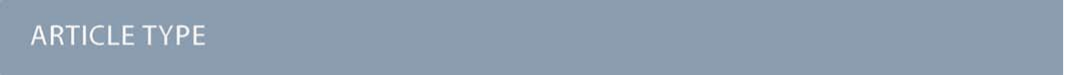}}\par
\vspace{1em}
\sffamily
\begin{tabular}{m{4.5cm} p{13.5cm} }

\includegraphics{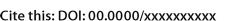} & \noindent\LARGE{\textbf{Efficacy of simple continuum models for diverse  granular intrusions}} \\
\vspace{0.3cm} & \vspace{0.3cm} \\

 & \noindent\large{Shashank Agarwal\textit{$^{a}$}, Andras Karsai\textit{$^{b}$}, Daniel I Goldman\textit{$^{b}$}, Ken Kamrin\textit{$^{a \ast}$}} \\

\includegraphics{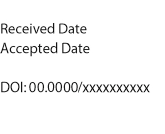} & \noindent\normalsize{Granular intrusion is commonly observed in natural and human-made settings. Unlike typical solids and fluids, granular media can simultaneously display fluid-like and solid-like characteristics in a variety of intrusion scenarios. This multi-phase behavior increases the difficulty of accurately modeling these and other yielding (or flowable) materials. Micro-scale modeling methods, such as DEM (Discrete Element Method), capture this behavior by modeling the media at the grain scale, but there is often interest in the macro-scale characterizations of such systems. We examine the efficacy of a macro-scale continuum approach in modeling and understanding the physics and macroscopic phenomena in a variety of granular intrusion cases using two basic frictional yielding constitutive models. We compare predicted granular force response and material flow to experimental data in four quasi-2D intrusion cases: (1) depth-dependent force response in horizontal submerged-intruder motion; (2) separation-dependent drag variation in parallel-plate vertical-intrusion; (3) initial-density-dependent drag fluctuations in free surface plowing, and (4) flow zone development during vertical plate intrusions in under-compacted granular media. Our continuum modeling approach captures the flow process and drag forces while providing key meso- and macro-scopic insights. The modeling results are then compared to experimental data. Our study highlights how continuum modeling approaches provide an alternative for efficient modeling as well as a conceptual understanding of various granular intrusion phenomena. 

} \\
\end{tabular}
 \end{@twocolumnfalse} \vspace{0.6cm}
 ]

\renewcommand*\rmdefault{bch}\normalfont\upshape
\rmfamily
\section*{}
\vspace{-1cm}


\footnotetext{\textit{$^{a}$~77 Massachusetts Ave, Cambridge, USA. }}
\footnotetext{\textit{$^{b}$~North Ave NW, Atlanta, USA.}}
\footnotetext{*\, E-mail: kkamrin@mit.edu}






\section{Introduction}
Granular intrusion is a common scenario in various natural and human-made situations \cite{soliman1976effect,omidvar2014response,ortiz2019soft}
. In many cases, the flow response to intrusion leads to a multi-phase behavior of the granular media, where the media shifts among solid-like, fluid-like, and gas-like states \cite{van2017impact,andreotti2013granular}. This behavior depends on material properties and local state variables such as pressure and granular packing fraction\cite{wood1990soil}. The space and time dependence of these state variables, as well as the multi-phase behavior of the media, makes granular flow response challenging and computationally expensive to model.

Discrete Element Methods (DEM) are reliable and established methods for modeling granular media \cite{jensen1999simulation,jensen2001simulation}. In DEM, the granular medium is modeled at the particle level, with the frictional, elastic contact forces between adjacent and interacting particles obtained using standard contact models like Hertzian and Hookean contacts\cite{brilliantov1996model,silbert2001granular}, generally with an assumption of no plastic yielding of the individual grains. The elastic and external forces are integrated via Newton's laws of motion to calculate particle velocities and positions in DEM. In the past few decades, DEM has played a major role in the advancement of the field of granular media\cite{kamrin2012nonlocal,kim2020power}. However, the precision of DEM comes at a computational cost. Once the simulation's spatial scale becomes large relative to the size of the particles in the system, the computational cost of simulating many particles makes DEM prohibitely expensive. As an example, a 0.1 mm particle diameter granular system with a 0.5m per side cubic domain would require evaluating $\sim$ $9\times 10^{11}$ DOF per time step (at a packing fraction of 0.64 with 6 degrees of freedom, DOF per particle). While current large-scale granular DEM studies remain in the range of $\sim10^7$ particles ($\sim10^8$ DOF) \cite{jajcevic2013large,zhong2016cfd}, efforts are being made to extend this range. Most recently, some have reached the $\sim10^9$ DOF range with the use of supercomputers and GPUs \cite{kelly2019billion}.\par 

In recent years, there has been a growing interest in modeling increasingly large granular systems, such as those relevant in ballistic impacts, wheeled locomotion, and agricultural plowing \cite{soliman1976effect,omidvar2014response,ortiz2019soft,vriend2013high,daniels2008force}. This has led to a push for computationally faster alternatives to DEM, such as using continuum descriptions of granular media and further reduced models such as Resistive Force Theory \cite{agarwal2019modeling,li2013terradynamics,agarwal2021surprising}. At the fundamental level, cohesionless granular media can be described as a plastic material with a frictional yield criterion and no cohesion strength (hence no ability to support tension). 
But, a comprehensive continuum model for fully describing a granular material requires incorporating a variety of behaviors shown by the media arising from the constituent grain shape, size, stiffness, distribution, and inter-particle interaction properties\cite{andreotti2013granular}, as well as effects due to the surrounding air (pressure and flow-drag) on the granular media \cite{mahabadi2017impact}. Thus, often scenario-specific simplifications are made during constitutive model development. Conversely, even for a simple granular material, a constitutive framework encapsulating all granular phenomena would have to merge many different pieces together, e.g. dissipative kinetic theory \cite{brilliantov2010kinetic,jenkins1985kinetic} in dilute regimes; strengthening based on packing fraction \cite{schofield1968critical} and internal state variables (like fabric and hysteresis) evolution \cite{yu2008non}; rate-dependence of the strength (such as $\mu(I)$ rheology \cite{jop2006constitutive}); non-trivial yield surface shapes in stress space \cite{wiacek2011mechanical}; and particle size effects \cite{henann2013predictive,kamrin2019non} in the dense flow regime. 
\par

A variety of computational methods have been developed and utilized recently to aid in simulating continuum models for large deformation processes, such as the Material Point Method (MPM){\cite{dunatunga2015continuum,agarwal2019modeling,daviet2016semi,daviet2016nonsmooth}}, the Particle Finite Element Method (PFEM)\cite{davalos2015numerical}, and Smoothed Particle Hydrodynamics (SPH)\cite{peng2016unified}. In this regard, a recent study  on high-speed locomotion of wheels \cite{agarwal2021surprising} in grains, a process which involves complex trans-phase  characteristics of the soil and non-trivial grain motions, revealed that a continuum treatment implemented with MPM captures the essential behaviors and agrees well with experimental data.   

In the present work, we highlight the capability of the continuum approach in modeling and explaining diverse granular intrusion phenomena using a very minimal set of constitutive ingredients. The phenomena we focus on herein have been independently studied in the literature, and our goal is to demonstrate that they can be quantitatively modeled and unified under a family of basic constitutive assumptions. We first introduce two constitutive representations of non-cohesive granular media: a non-dilatant plasticity model (NDPM), and a dilatant plasticity model (DPM). {The DPM model permits dilatancy during dense flow while the NDPM model assumes dense flow is a constant-volume process (details later). Thus,} the NDPM is suitable for modeling steady-state granular behaviors, while the DPM is a more suitable model for transient flow processes. The latter is also a more elaborate model that converges to the former model under certain limits. Both of these models have their respective advantages depending on the scenarios they simulate. We do not include micro-inertial $\mu(I)$ effects \cite{yu2008non} (effects of grain-level inertia on material properties) in these models; thus both of these constitutive models are rate-insensitive.  

\begin{figure}[h]
\centering
 \includegraphics[trim = 0mm 95mm 200mm 0mm, clip, width=1.0 \linewidth]{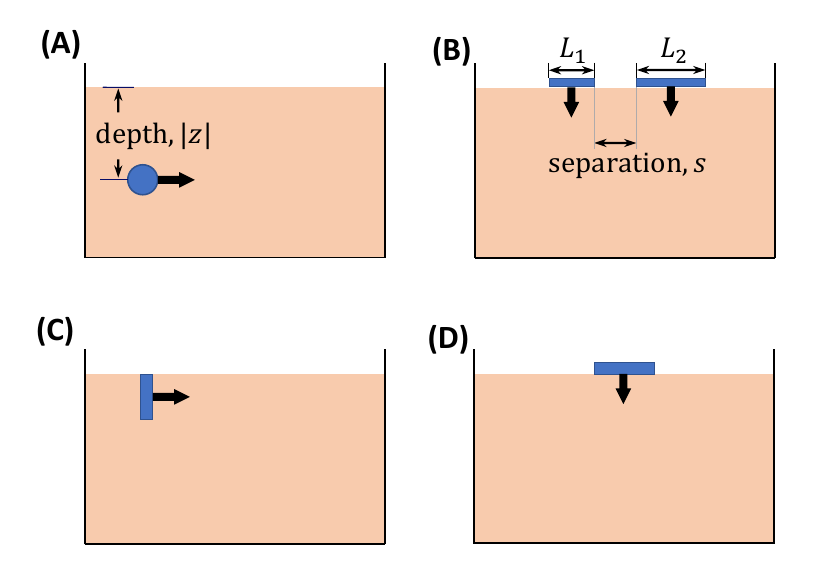}
 \caption{\emph{Schamatics of four test cases considered in this study:} \emph{(A)} Horizontal dragging of a submerged cylinder at different depths, \emph{(B)} multiple-plate vertical granular intrusion, \emph{(C)} free-surface plowing in under- and over-compacted granular media, and \emph{(D)} vertical intrusion in under- and over-compacted granular media.  Cases \emph{(A)} and \emph{(B)} are be modeled with the non-dilatant plasticity model (NDPM), and cases \emph{(C)} and \emph{(D)} are modeled with the dilatant plasticity model (DPM) to model transient effects.
 }
 \label{fig:1}
\end{figure}

We demonstrate the utility of each model by considering granular flow and force responses in four fundamental intrusion cases which have been studied in the literature: (1) depth-dependent force response in horizontal submerged intruder motion; (2) separation-dependent drag variation in parallel plate vertical intrusion; (3) initial density-dependent drag fluctuations in free surface plowing; and (4) flow zone development during vertical plate intrusions in under-compacted granular media (Figure \ref{fig:1}). We use NDPM in the first two cases (which focus on a steady response), and DPM in the latter two cases (where transient effects are the focus). Our continuum modeling approach captures the dynamics of such granular flows and gives a deeper macroscopic insight into each of these cases. Thus, our study highlights the efficacy of continuum modeling for predictive purposes, as well as in developing macroscopic conceptual understanding of diverse granular intrusion phenomena.

\par

\section{Material and Methodology}
We use two constitutive models to represent granular volumes in this study namely: the non-dilatant plasticity model (NDPM), and the dilatant plasticity model (DPM). We first explain the two models in detail and then explain the numerical method we use to implement the schemes. 

\textbf{Non-dilatant plasticity model (NDPM):
} This model is taken from Dunatunga and Kamrin \cite{dunatunga2015continuum,dunatunga2017continuum}. The model assumes a rate-insensitive, Drucker-Prager yield criterion, incompressible plastic shear flow (with no dilatancy), and cohesionless response in extension whereby the material becomes stress-free below a \textit{`critical density'}. The  constitutive flow equations representing the above behavior are shown below. These simultaneous constraints describe the material's separation behavior, shear yield condition, and tensorial co-directionality, respectively:
\begin{align}
&(\rho-\rho_d)P=0 \quad \text{and} \quad P\ge 0 \quad \text{and} \quad \rho\le\rho_d, \label{eq:1}
\\
&\dot{\gamma}({\tau}-\mu_s P)=0 \quad \text{and} \quad \dot{\gamma}\ge 0 \quad \text{and} \quad {\tau}\leq\mu_{s}P, \label{eq:2}
\\
&D_{ij}/\dot{\gamma} = \sigma_{ij}'/2{\tau} \quad \text{if} \quad \dot{\gamma}>0 \quad \text{and} \quad P>0, \label{eq:3} 
\end{align}

for $i,j=1,2,3$. \\
where, \\
$\rho $ \quad is the local granular density, \\
$\rho_d $ \quad is the critical density of close-packed grains, \\
$\mu_s$ \quad is the internal friction coefficient,\\
$P = -{\sigma}_{ii}/3$ \quad is the local hydrostatic pressure, \\
$\dot{\gamma}=\sqrt{2 D_{ij}D_{ij}}$ \quad is the equivalent shear rate. \\
${\tau}$ = $\sqrt{\sigma'_{ij}\sigma'_{ij}/2}$ \quad is the equivalent shear stress, \\
${\sigma}'_{ij}$= ${\sigma}_{ij}$ + $P\delta_{ij}$ \quad is the deviatoric part of Cauchy stress tensor,\\
$D_{ij}=(\partial_iv_j+\partial_j v_i)/2$ \quad is the flow rate tensor. \par

For simplicity, the equations above are expressed in rigid-plastic form, where the flow rate is approximately all due to plastic flow.  However, we also include a small elastic component which ensures that below the yield criterion the grainshave a well-defined solid constituive behavior. The model evolves the flow by solving the momentum balance equations, $\partial_{j}\sigma_{ij}+\rho g_i=\rho \dot{v}_i$.  We also assume a constant friction coefficient {$\mu_f$} between the granular continuum and solid-body surfaces. {The internal friction value $\mu_s$ is often measured from the tangent of the angle of repose for the granular media or from a direct shear test\cite{taylor1948fundamentals}}. We refer to this model as the `Non-Dilatant Plasticity Model' (NDPM). We provide the input material properties used in various cases in the relevant sections. A basic implementation of this continuum modeling approach can be downloaded for Matlab \cite{agarwal2020}.

\textbf{Dilatant plasticity model (DPM):
}While NDPM assumes a single critical granular density cutoff for the entire granular volume in the system, this is often not the case in real granular systems. Granular media can support stress in a range of densities including close and loose packings\cite{andreotti2013granular}. The second model we present addresses this limitation of the NDPM model and allows for density evolution in the media \cite{wood1990soil}. This model takes inspiration from the family of Critical State models \cite{wood1990soil,schofield1968critical}. We use a simple rigid-particle version of the Critical State model in which grains are not modeled to crush under hydrostatic loads; rather, they only have a hydrostatic elastic response in this case. When pressure is non-zero,  the material is deemed to be in a `dense state' and the density evolution occurs through shearing-based Reynolds dilation/contraction, representing particle rearrangements at the granular level. In a dense state, the packing fraction $\phi$ is equal to an evolving state variable $\phi_d$. Every shearing action in dense media drives the local value of $\phi_d$ towards a limiting steady-state critical packing fraction, $\phi_c$. Furthermore, when the packing fraction varies, the material's friction coefficient also changes; material with low packing fraction has low consolidation and thus has less strength than more consolidated media. \par
Thus, in terms of constitutive relations, the simultaneous constraints describing the material’s separation behavior and shear yield condition (eq.\ref{eq:1} and eq.\ref{eq:2}, respectively) remain the same with the two exceptions that the close-packed density, $\rho_d$, and internal friction value, $\mu_s$, are non-constant local state variables (which vary with time and space), as follows:  
\begin{align}
& \rho_d = \rho_g \phi_d,    & \mu_s = \mu_{c} + (\phi_d-\phi_c)\chi \label{eq:4}
\end{align}
Here, $\rho_g$ represents the solid grain density, $\chi$ represents a dimensionless scaling constant, and $\mu_c$ represents the critical-state internal friction value reached by material at steady state (when $\phi_d= \phi_c$). When pressure is positive (the packing is dense), the evolution of $\phi_d$ is modeled as: 
\begin{align}
&\frac{d\phi_d}{dt} = -3\beta \phi_c  \dot{\gamma}  \label{eq:5}
\end{align}
where, $\beta = (1/3) (\phi_d-\phi_c)\chi$ is a local dilatancy variable. In the absence of {confining} pressue, the material can leave the dense state ($\phi<\phi_d$), but reconsolidates at the current value of $\phi_d$ unless the material opens up beyond a global lower limit, $\phi_d^{\textrm{min}}$, which defines a minimum density possible for loaded media, below which no connected material states exist. We set this value to $0.45$ in all simulations. Thus, the material can exist in a dense state only for $\phi \ge 0.45$ and the local state variable $\phi_d$ gets reset to $\phi_d^{\textrm{min}} (= 0.45)$ whenever $\phi$ drops below this value.

In addition to the material’s separation behavior and shear yield condition (eq.\ref{eq:1} and eq. \ref{eq:2}), the material flow rule has a deviatoric dependence as before (eq.\ref{eq:3}) but due to dilation now also has a corresponding volumetric component. The volumetric strain-rate is given by the dilatancy variable, $\beta$. The plastic flow rate tensor, thus, has the form:  
\begin{align}
&\boldsymbol{D}_{ij}/\dot\gamma = \boldsymbol{\sigma'_{ij}}/2\tau + \beta\delta_{ij}
\quad \text{if} \quad \dot{\gamma}>0 \quad \text{and} \quad P>0 \label{eq:6}
\end{align}
Thus, the combination of eqs.\ref{eq:1}, \ref{eq:2}, \ref{eq:4}, \ref{eq:5}, and \ref{eq:6} collectively represent  the Dilatant Plasticity Model (DPM). Note that DPM reduces to NDPM in the limit of $\chi \rightarrow 0$. {Regarding its uasge in cases 3 and 4, $\chi$ is calibrated for each material qualiatively based on experimental observations.}

While DPM is a more elaborate representation of granular media compared to NDPM, NDPM is easier to implement and is computationally more efficient to solve. In this study, we aim to highlight the advantages and sufficiency of using each of these constitutive forms on a case-by-case basis. 

\begin{figure}[h]
\centering
 \includegraphics[trim = 0mm 130mm 150mm 0mm, clip, width=1.0 \linewidth]{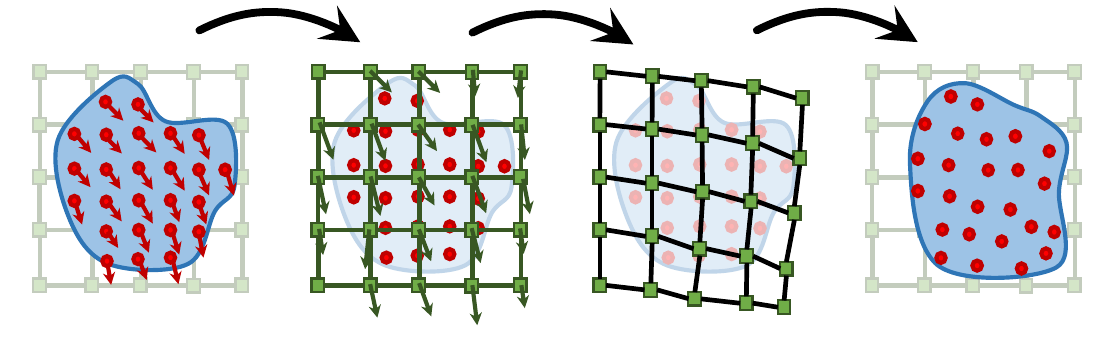}
 \caption{\emph{Sample explicit time integration step in MPM:} A sample explicit time-integration step used in an MPM implementation. The Lagrangian tracers representing material points (solid red circles) carry the state of material over time and space. The background grid (green squares) assists in integrating the motion on the simulation domain and is reset each step. More details can be found in Dunatunga and Kamrin \cite{dunatunga2015continuum}.}
 \label{fig:2}
\end{figure}

For the numerical implementation, we use a continuum simulation approach based on the Material Point Method (MPM)\cite{sulsky1994particle}, a derivative of the fluid-implicit-particle (FLIP) method for implementing a continuum description of granular media. The MPM implementation uses a combination of Lagrangian material points, which contain the state of the continuum, and a background computational mesh that is used to solve the equations of motion. Since the state is saved on the material points, the mesh is reset at the beginning of each computational step without any information loss, and thus the method allows for large deformations in the media without the error associated with large mesh deformations. A schematic representation of an explicit time-integration MPM-step is shown in Figure \ref{fig:2}. We use the MPM implementation described in Dunatunga and Kamrin \cite{dunatunga2015continuum} to implement the different constitutive equations representing granular volumes assuming 2D plane-strain motion. 
{We choose the spatial grid resolution ($\Delta x$) in each system on a case-by-case basis, such that all the geometric features are well represented and all system outputs converge. Also, to ensure numerical stability, we choose a time step ($\Delta t$) that satisfies the CFL condition based on minimum element size, maximum Young's modulus, and minimum medium density.}\par

\section{Results and Discussion}
\subsection{Drag and Lift on submerged cylinder dragging}
Drag and lift forces on submerged objects in granular media are relevant in processes such as mixing, mining, soil-buried pipelines, and animal locomotion \cite{jarray2019cohesion,aguilar2016review,li2019review}. Over the past two decades, several studies\cite{zhang2015numerical,guillard2014lift,wu2020shallow,ding2011drag} have explored the mechanisms and the variation of these forces  by considering the horizontal dragging of submerged, rigid cylinders (at different depths) as a test case.  {We specifically consider the work of Guillard et al.\cite{guillard2014lift}, who find that the horizontal drag force on cylindrical objects moving horizontally increases with increasing depth while the vertical lift force plateaus for depths greater than a O(1) factor of the cylindrical diameter.}
The observations consider a depth range deep enough so that the cylinders are completely immersed in the granular bed even at the minimum depth. At very low depths near the free surface, the dragging motion of the cylinders results in an accumulation of the media in front of the intruders. This accumulation can augment the depth-dependence of drag forces and thus can change the force trends mentioned above. Thus, we do not consider such a near free-surface depth range in this study {(as in Ding et al.\cite{ding2011drag})} and focus on depths$\ge 2.5D$. \par

\begin{figure*}[h!]
\centering
 \includegraphics[trim = 0mm 0mm 10mm 0mm, clip, width=1.0 \linewidth]{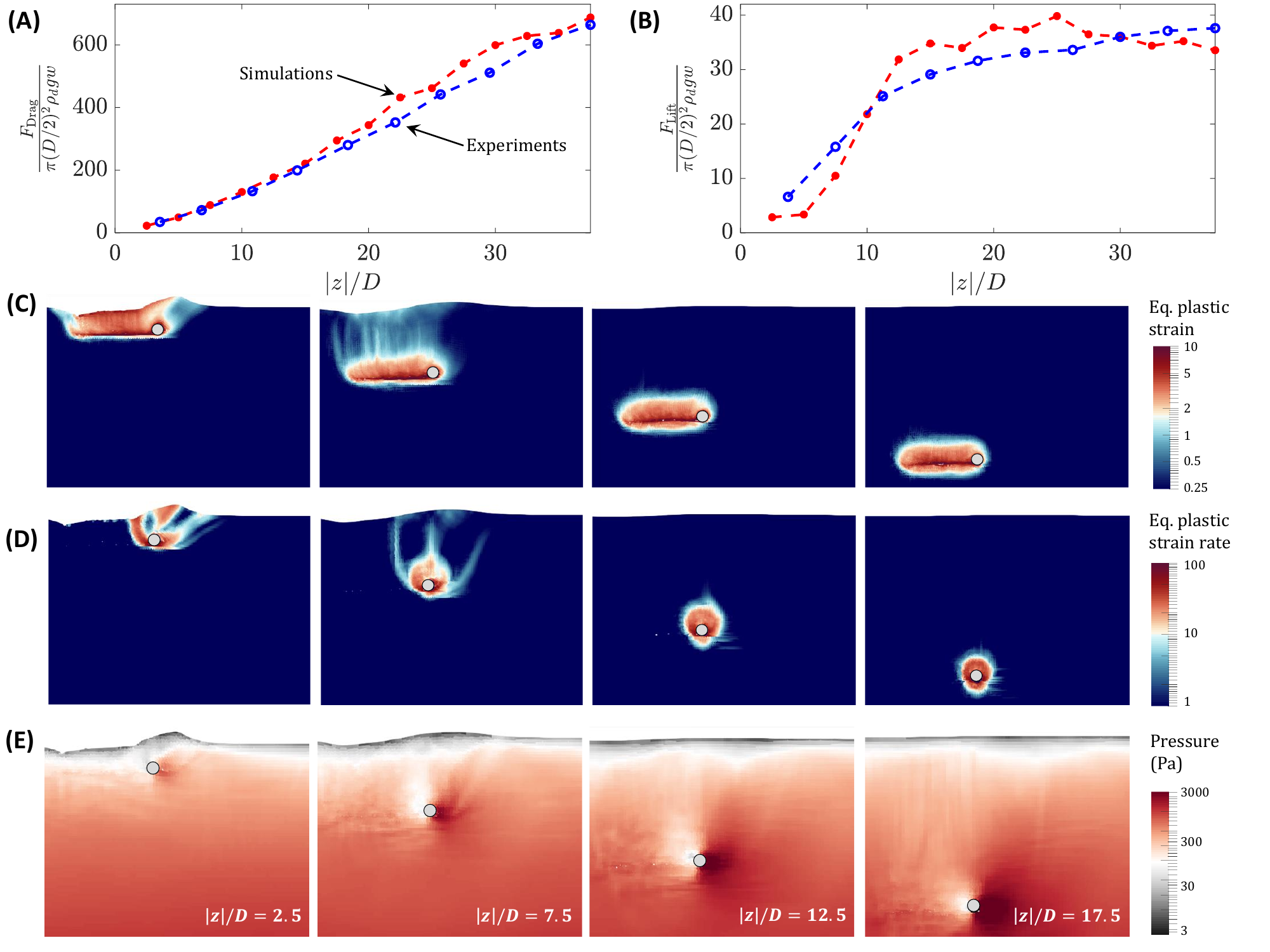}
 \caption{\emph{Case 1: Experimental vs Simulations results for drag and lift on submerged cylinders {(using NDPM model)}}: Comparision of experimental data \cite{guillard2014lift} (in blue) and the continuum simulations results (in red) for variation of \emph{(A)} drag, and \emph{(B)} lift on submerged cylinder motion in granular media with depth, $|z|$. The drag and lift forces are non-dimensionalized with characteristic force $\pi(D/2)^2\rho_d g w$ where $D$ is the cylinder diameter, $\rho_d$ is the effective medium density, $g$ is the gravity, and $L$ is out-of-plane cylinder width. Both experiments and continuum results use $D=4$ mm, $\rho_d=1512$ kg/m$^3$, and $g=9.8$ m/s$^2$. Similarly depth is non-dimensionalized with cylinder diameter, $D$. Variation of macroscopic state variables \emph{(C)} \textit{equivalent plastic strain}, \emph{(D)} \textit{equivalent plastic strain rate}{, and \emph{(E)} \textit{local hydrostatic pressure} } at four different depths, $|z|= [10,30,50,70]$ mm. The gray circle indicates the cylinder position and the drag direction is from left to right. See Movie S1 for visualizing material flow over time for cylinder drag cases considered in \emph{(C)} and \emph{(D)}.}
 \label{fig:3}
\end{figure*}

The experimental drag and lift results in Guillard et al.\cite{guillard2014lift} were obtained by evaluating torques and vertical lifts on horizontal cylinders inside large granular beds rotated about the vertical axis. 
After a half rotation or so about the vertical axis, the authors observed a reduction in the force response as the cylinder begins re-interacting with the disturbed material as it is rotated multiple times. 
Here, we use the quasi-steady data obtained from the first half rotation of the cylinder, before  it interacts with the disturbed media.  Presumably, density variations play a smaller role in this portion, and hence we choose to compare the behavior observed with that of the NDPM model.  \par

A schematic representation of the cylinder drag case is provided in Figure \ref{fig:1}A. For the NDPM implementation, we use a grain density $\rho_g$ of 2520 $\textrm{kg}/\textrm{m}^3$, and a critical packing fraction $\phi$ of 0.60. {We calibrate the internal friction value ($\mu_s$) for glass beads to $0.48$ to accurately model the initial slope of depth vs drag graph from the experiments. This is inline with the expected range of $\mu_s$ for the glass beads which lies in the $0.39-0.55$ range (corresponding to an angle of repose range $\sim22^{\circ}-29^{\circ}$, depending on the surface roughness) \cite{klinkmuller2016properties}}. The calibration absorbs possible inconsistencies between 2D simulations with 3D experiments and also incorporates the variations due to indirect measurements of experimental drag and lift from rotating cylinder experiments\cite{guillard2014lift}.

The cylinder is modeled as an elastic body with a high elastic modulus, thereby acting as an approximate rigid body. The media-cylinder interface friction coefficient ($\mu_f$) is set to $0.35$. The cylinder diameter, $D$ is $4$ mm, and the presumed out-of-plane length, $w$ is $1$ m (as the simulations are 2D plane-strain). {We use a $0.20$ m $\times$ $0.16$ m granular bed, and a $5\times10^{-4}$ m spatial resolution ($\Delta x$) for simulating these cases.}  

Figure \ref{fig:3}A and B show the comparison of Guillard's experimental data to our continuum results. The linearly increasing nature of drag vs depth graphs, along with the plateauing of lift vs depth graphs are well captured. The visualizations of plastic strains and strain rates (Figure \ref{fig:3}C and D) qualitatively agree with those reported in previous numerical and experimental studies \cite{zhang2015numerical,guillard2014lift}. {Notably, the variation of the equivalent plastic strain shows a strong semblance with the work of Wu et al.\cite{wu2020shallow}. Additionally, we also plot the variation of the local hydrostatic pressure at four different depths in figure \ref{fig:3}E. The pressure distribution shows an asymmetry about the intruder, consistent with the existence of both drag and lift net forces, and as the intruder's depth increases the vertical asymmetry diminishes, which is consistent with the plateauing of the lift force and continued growth of drag force.}

Similar to past DEM studies by Guillard et al.\cite{guillard2014lift} and experimental studies by Wu et al.\cite{wu2020shallow}, Figure \ref{fig:3}C and D show high localization of the flow field for the depths greater than the lift turnover depth. A further increase in depth after this point results in a minimal-to-no change in the material flow profile. The drag and lift force behaviors are also related to this flow localization behavior. While the drag forces continue to increase linearly with depth, the lift forces become saturated near the depth the flow localizes, indicating a possible correlation between the flow and force. {Figure \ref{fig:3}E indicates a reducing asymmetry of the pressure variation around the cylinder as depth increases, consistent with the notion that drag forces continue to increase but lift forces plateau}. A deeper analysis of this variation {(i.e. the distribution and the relative magnitude of pressure) to understand }the mechanics of the force distributions on different faces of the intruders is reserved for future study. 

Thus, the case of dragging submerged cylinders provides an example where NDPM captures the observed intrusion behavior. 

\subsection{Vertical drag in two-plate granular intrusions}

Multi-body intrusions offer another commonly encountered scenario in various real-life situations. Several researchers \cite{de2016lift,merceron2018cooperative,pravin2020effect} have examined the dynamics of granular intrusions involving multiple intruding bodies. The case we study specifically takes inspiration from the work of Swapnil et al.\cite{pravin2020effect} who examined the variation of vertical drag forces on a pair of parallel rigid plates during vertical downward intrusions as a function of the separation between them. They observed a peak in the average vertical drag as the separation between the plates was increased. {A similar `cooperative effect' was observed by Merceron et al.\cite{merceron2018cooperative} during the upward motion of parallel intruders in the granular media. In their study separations below a critical distance between the intruders were found to jam the media between the plates (intruders) and result in a peak in the average particle disturbance (measured as average avalanche length) near this critical separation between the intruders.} 
We use NDPM to investigate the findings of Swapnil et al.\cite{pravin2020effect} and to capture and obtain macroscopic insight into the phenomenon. We also conduct experiments to verify claims that follow from theoretical analysis of the NDPM model. \par

\begin{figure*}[h]
\centering
 \includegraphics[trim = 0mm 25mm 5mm 0mm, clip, width=1.0 \linewidth]{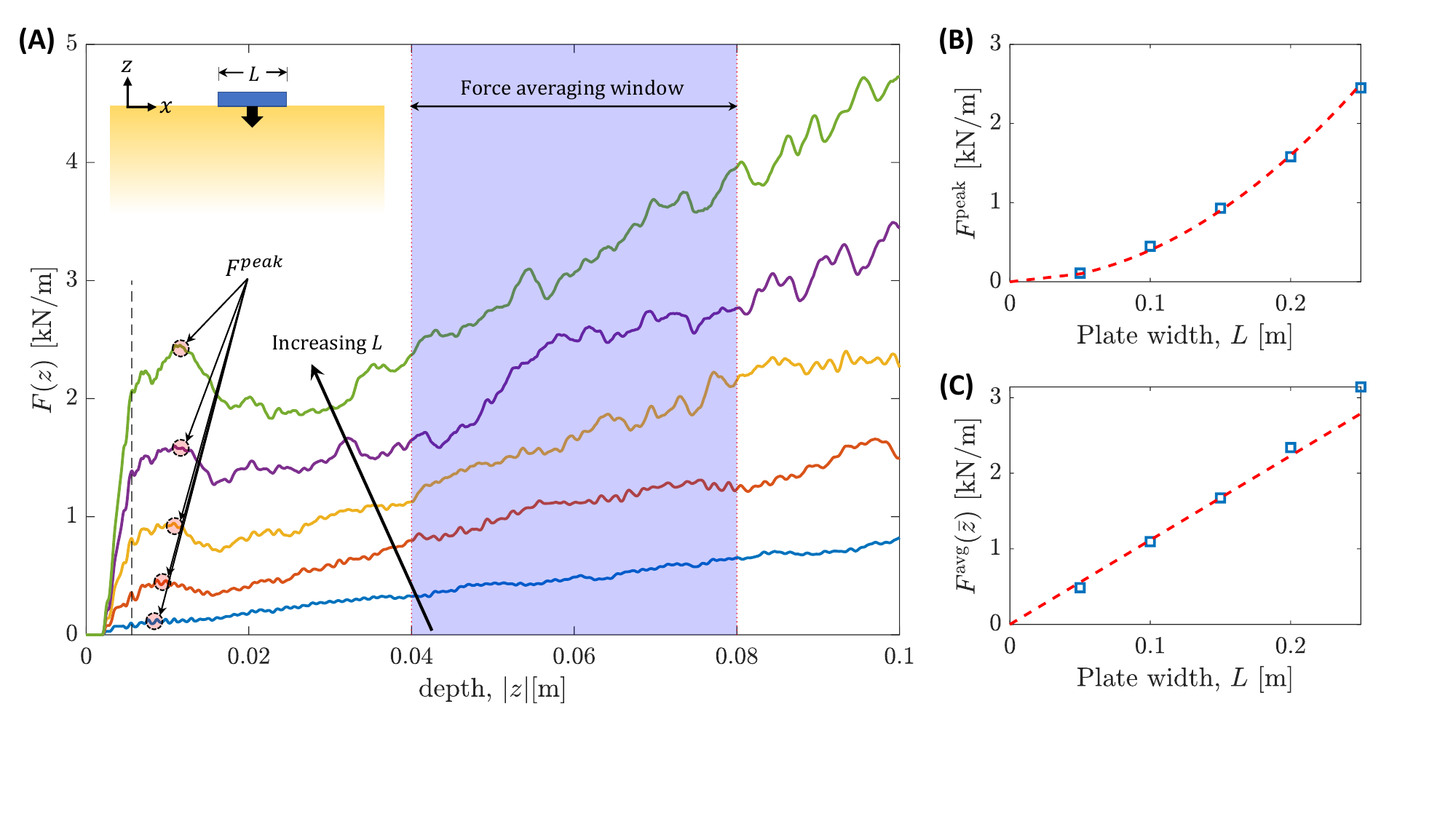}
\caption{\emph{Case 2: Vertical intrusion of single plates} {(using NDPM model)}: \emph{(A)} Variation of drag forces with intrusion depth for plates of various widths, $L$ $(0.05, 0.10, 0.15, 0.20, 0.25 \, \textrm{m})$ during single-plate vertical granular intrusion. The arrow shows the direction of increasing $L$. The simulations are plane-strain and glass bead properties are used for simulating granular media ($\rho_g$=2520 kg/m$^3$, $\phi=0.6$, $\mu_s=0.4$). Each data series on the left graph was time-averaged over a $0.5$ ms window to remove high-frequency force fluctuations. The vertical black dotted line in \emph{(A)} shows the depths after which this averaging window includes sufficient data. Dependence of \emph{(B)} initial force peaks, and \emph{(C)} average drag forces on plate widths in single plate intrusions (blue squares). The corresponding quadratic and linear fits for \emph{(B)} and \emph{(C)} are shown as red dotted lines in corresponding graphs. The force peaks are shown with arrows in the left graph, and the time window used for calculating the average forces is the highlighted blue region (\textit{force averaging window}) on the left graph. }
 \label{fig:4}
\end{figure*}

A schematic representation of the case is given in Figure \ref{fig:1}B. The case is studied in two parts. In the first part, we use generic material properties (provided next) to determine the behavior of the continuum model in the single-plate intrusion case. And in the second part, a direct quantitative comparison of simulation results with the experiments is done in the two-plate intrusion case (details later). We use a set of generic NDPM fitting properties from the earlier case, i.e. a grain density, $\rho_g$ of 2520 $\textrm{kg}/\textrm{m}^3$, a critical packing fraction $\phi$ of 0.6, but choose the internal internal friction coefficient to be $\mu_s=0.4$, which is common for glass beads \cite{klinkmuller2016properties}.
The plates were modeled as stiff elastic bodies with vertical displacement of control points assigned; thus, the plates act as quasi-rigid objects with a common fixed downward velocity. The intrusion velocity was set to $0.1$ m/s in all simulations, and the media-plate surface friction was set to $0.35$. {We use a $1.2$ m $\times$ $0.6$ m granular bed, and a $1.25\times10^{-3}$ m spatial resolution ($\Delta x$) for simulating these cases.}

\subsubsection*{Understading single plate intrusions }
We begin by first analyzing the variations of vertical drag forces in single plate intrusions. Figure \ref{fig:4}A shows the drag forces on vertical plate intrusions under NDPM for various plate widths at a constant intrusion velocity of $0.1$ m/s. {Note that the same model and implementation was used previously \cite{dunatunga2017continuum} in the related case of high-speed impacts of a circular intruder in granular media; the model quantitaively matched the flow and force data of Clark and Behringer \cite{clark2013granular}}. Before analyzing the simulation results, we perform a scaling analysis of the problem assuming the NDPM equations hold, in order to predict the dependence of drag on various system parameters. In this case, we expect the drag force, on a vertical intruder of width $L$ and out-of-plane width $w$ at a depth of $z$, to depend on various system parameters, i.e. intruder dimensions ($w$ and $L$), depth $z$(distance between the intruder bottom and the free-surface), close-packed media density $\rho_d$, gravity $g$, the media's internal friction $\mu_s$, the media-intruder interface friction $\mu_{f}$, and the velocity of intrusion $v$. Using scaling analysis, with base units of length as $L$, time as $\sqrt{L/g}$, and mass as $\rho_d L^3$, we obtain: 
\begin{align}
F^{D} &= \rho_d L^3 L (\sqrt{L/g})^{-2} \hat{f}(\mu_s,\mu_f,z/L,w/L, v/\sqrt{gL}) \nonumber\\
 &= \rho_d g L^3 \hat{f}(\mu_s,\mu_f,z/L,w/L, v/\sqrt{gL}) \label{eq:7}
\end{align}
{Where, $\hat{f}$ represents some unknown function. Recall that we are scaling based on NDPM, so certain properties such as the particle diameter do not appear in the above relation.} \par

The dependence of $F^D$ on $z$ can be divided into two regimes. (1) At low depths $ z \ll L $, the variable $z/L$ is negligible and can be ignored. And (2), at larger depths, the drag forces $F^{D}$ are known to show a linear dependence on depth (after an initial jump in the vertical drag near free surfaces\cite{PhysRevLett.126.218001,katsuragi2007unified,brzinski2013depth,kang2018archimedes}). In both of these regimes, we  set the dependence of $F^{D}$ on $w$ to be linear assuming the plane-strain nature of the intrusions. We also focus on intrusions that are sufficiently slow such that $F^{D}$ is independent of velocity $v$ (low-velocity regimes) as seen in the work of Swapnil et al.\cite{pravin2020effect}. \par
With the above assumptions, in the $z \ll L$ regime, we obtain: 
\begin{align}
&F^{D} = \rho_d g L^3 (w/L) \hat{f}(\mu,\mu_f) \nonumber \\
&\to F^{D}/w = \rho_d g L^2 \hat{f}(\mu,\mu_f) \nonumber \\ 
&\to {F} \propto L^2. \label{eq:8}
 \end{align}
And in the latter, moderate-depth regimes with $F^D \propto z$, we obtain: 
\begin{align}
&F^{D} = \rho_d g L^3 (w/L) (z/L) \hat{f}(\mu,\mu_f) \nonumber\\
&\to F^{D}/w = \rho_d g z L \hat{f}(\mu,\mu_f) \nonumber\\
&\to {F} \propto z L. \label{eq:9}
\end{align}
Thus, from the scaling analysis, we expect the drag forces per unit out-of-plane width ${F}\ (=F^{D}/w)$ (unit $N/m$) to have a quadratic dependence on plate width, $L$, near the free surface, and a linear dependence on plate width at larger depths. {The scaling analysis does not provide information on the exact form of variation of  $F^D$ in $z$ in the region connecting the two regimes. However, we do expect the variation to be non-monotonic in $z$ since in the vanishing depth limit the force scales as $L^2$ and in the deeper limit it scales with $zL$.}  \par

Using the simulation's force output, in Figure \ref{fig:4}B and C we plot the observed relationships between $F$ and $L$ at, respectively, a small{er} depth (where initial peaks occur) and in a deeper regime, where linear force variations are predicted.  The forces in Figure \ref{fig:4}C are averaged over a depth window {$(z_i-z_f)$}, i.e.  $F^{avg} = (\int_{z_i}^{z_f}{F(z)\text{d}{z}})/(z_f-z_i)$. The expected trends from the dimensional analysis are apparent. Besides our own simulation results, the linear dependence of drag force on intruder area, once deep enough,  is a well studied relation \cite{PhysRevLett.126.218001,katsuragi2007unified,brzinski2013depth}. 
We reiterate that our simulations are all in the quasistatic regime. Thus, the velocity contributions are negligible in comparison to static force contributions in our study. {Thus, the initial peak in the force response is not related to inertial drag as seen in faster intrusions\cite{bester2017collisional,katsuragi2007unified}}.  

Physically, the initial force maxima ($F^{peak}$) in the force vs. displacement graphs of Fig. \ref{fig:4}A correspond to the force requirements for initiating media flow in the system. The $F^{peak}$ force contributions enable the beginning of flow by initiating the shearing of the media, which exists at some finite strength due to finite pressure under the plate. From slip-line theory for granular intrusions\cite{sokolovskiui1960statics}, the flow develops a wedge-shaped no-shear-zone below the intruder, which has an acute angle of $\pi/2 - \tan^{-1}(\mu)$ (where $\mu$ represents the material internal friction). Thus, the requirement of shearing media of a finite strength along the wedge-shaped shear zone is responsible for these force contributions. We can obtain an intuition for the initial $L^2$ dependence of force by considering the conventional limit analysis for indentation of materials with yield stress $Y$, common in manufacturing processes such as drawing and blanking\cite{bower2009applied}. In such cases, indenter force per unit out-of-plane width, $F$, varies as $F\propto Y\times L$ for $L$ the indenter plate length. In frictional media, the strength is pressure-sensitive. With $Y = \mu \times P$ and supposing $P$ grows linearly along the edge of the no-shear-zone wedge, the mean strength along the edge grows $\propto L$. Substitution of $Y\propto L$ in the limit analysis formula then gives the observed quadratic dependence of the drag force on $L$. We do not further investigate these forces but identify that the existence of small regions of under-compacted granular media near free surfaces is expected to suppress the growth of such forces in many cases (see figure \ref{fig:8}D). Our simulations indicate that the drag forces' dependence of $F$ on $L$ enter linear regimes at depths $z$ $\sim$ $O(10^{-1})L$. {These forces are unique from the `added mass' effects\cite{aguilar2016robophysical} and other macro-inertial effects\cite{PhysRevLett.126.218001, katsuragi2007unified} in granular impacts that are common in high-speed intrusion (and vary $\propto v^2$) since our intrusion velocities are small.} 

 \par

\subsubsection*{Two plate intrusions}

\begin{figure*}[ht!]
\centering
 \includegraphics[trim =0mm 15mm 0mm 0mm, clip, width=1.0 \linewidth]{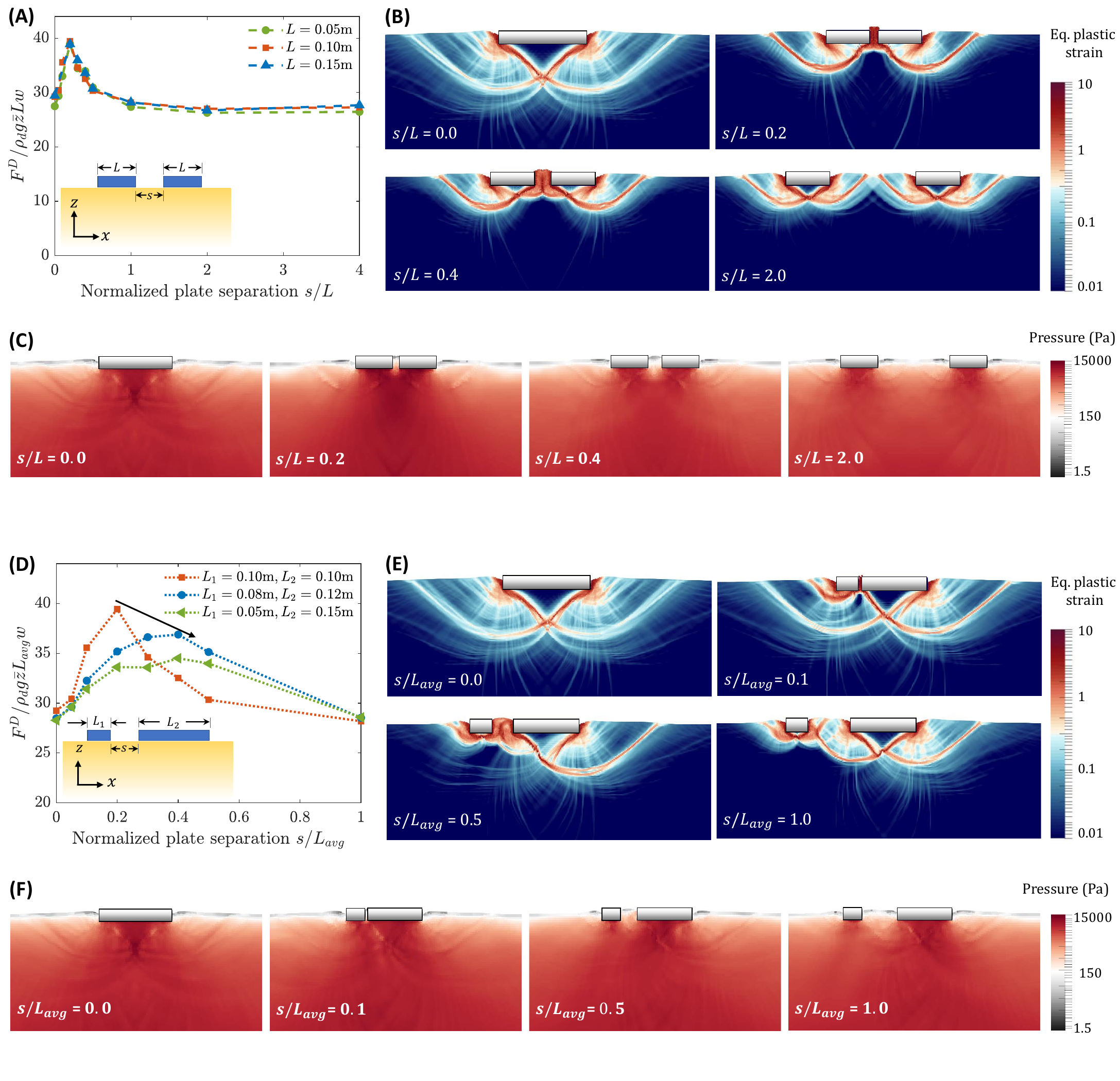}
 \caption{ { \emph{Case 2: Vertical intrusion of two parallel plates {(using NDPM model)}}: \emph{(A)} Variation of vertical drag (normalized) with plate separation, $s$ (normalised to plate width $L$) and \emph{(B)} the correspondng flow fields of \textit{equivalent plastic strain} for equally sized plates (L = 0.10 m each) and \emph{(C)} \textit{local hydrostatic pressure}, for equally sized ($L=0.10$ m each) plates. \emph{(D)} Variation of vertical drag with plate separation, $s$ (normalised with average plate width $L_{avg}=(L_1+L_2)/2$) and the flow fields visualization using \emph{(E)} \textit{equivalent plastic strain} and \emph{(F)} \textit{local hydrostatic pressure} for unequally sized plates ($L_1=0.05$ m and $L_2=0.15$ m, bottom) plates. The forces $F^D$ in \emph{(A)} and \emph{(D)} are averaged over a depth range $(z_i-z_f)$of $0.04-0.08$ m for all the cases and an average depth, $\bar{z} = 0.06$ m, $w=1$ m, $g=9.8 m/s^2$, $\rho_d= 1512 kg/m^3 (=\rho_g*\phi)$ is used. The simulations are plane-strain and glass bead properties are used for simulating granular media ($\rho_g$=2520 kg/m$^3$, $\phi=0.6$, $\mu_s=0.4$). See Movie S2 for visualizing material flow over time for the two cases considered in \emph{(B)} and \emph{(E)}.}}
 \label{fig:5}
\end{figure*}
The above scaling analysis gives an important insight into the case of multibody intrusions. In the purview of eq. \ref{eq:9}, we expect that two plates intruding far from each other and at moderate depth, will experience the same net vertical drag as that experienced by a single plate with an equivalent surface area --- the linearity in $L$ (eq.\ref{eq:9}) indicates that this equality should hold for any ratio of areas between the two plates. Thus, any combination of plate widths should experience the same forces both when they are infinitely far apart and when they have no separation. This analysis does not provide any information on the force variations when the plates are close but are at a finite distance from each other. However, physical intuition suggests that the presence of plates in the vicinity of each other will restrict the material flow, which should increase the drag on each plate. Given that the plates experience equal net forces at infinite and zero separation between them, we expect there to be a value of separation at which the force response is maximal or minimal (unless the force response is constant).  The work of Swapnil et al.\cite{pravin2020effect} explored this variation and found there is a peak in the force response at a low value of separation between the plates. Note that near the free surface ($z\ll L$), we do expect drag force at zero separation to be higher than at infinite separation due to the quadratic dependence of the force on plate size in this regime --- the ratio of forces for plate lengths of $L_1$ and $L_2$ would be  $F_0/F_{\infty} =(L_1+L_2)^2/(L_1^2+L_2^2)$ which is always greater than 1. 

Figure \ref{fig:5}A and D show the variation of force for different combinations of plate widths and plate separations. Continuum modeling shows the existence of force peaks for both equal plate cases (Figure \ref{fig:5}A) and unequal plate cases (Figure \ref{fig:5}D). Our observations are in accord with similar experiments and DEM simulations by Swapnil et al's\cite{pravin2020effect} for equal plates. 
 As the continuum modeling successfully captures the behavior, the detailed material states in these simulations can now be used to identify the macro-mechanical origins of the phenomena. We visualize the material flow by plotting snapshots of the plastic strain in a few of these cases. The plastic strain fields before, at, and after the force peak in Figure \ref{fig:5}B and E, suggest a macro-mechanical picture. We observe higher granular flow interaction between the two plates as the separation between the plates is decreased. For a single plate intruder, any neighboring flow restriction is expected to make it more difficult to push  material during the intrusion. Thus, decreasing plate separation results in increasing drag on each plate. Once the plates are sufficiently close, a large wedge-shaped rigid zone forms spanning the plates, causing the two plates to act as a large single plate. Thus, any further reduction in the separation does not result in additional flow restriction. Instead, it leads to a reduction in the effective area of the merged plate systems, and thereby the drag forces decrease upon a further decrease in separation. It is interesting to note that the material flow profiles when the two plates are together or far-separated are similar to the classical plasticity solution of Prandtl \cite{prandtl1920concerning} for yielding of metals upon indentation under a plate, characterized by a single rigid wedge of media under the indenter and flow emanating from both edges of the wedge. However, when the plates are separated but very close, the flow profile looks similar to the indentation plasticity solution of Hill \cite{hill1949plastic}, characterized by two smaller wedges under the plate and flow emanating only from the two outermost edges. {We also plot the variation of local hydrostatic pressure in the media for the two cases in figures \ref{fig:5}C and F. The ratio of local pressure magnitudes with equivalent hydrostatic pressure,  $\rho_d g|z|$, is large ($\sim 40$) in accord with the observations of Brezinski et al. \cite{brzinski2013depth}.} 

To understand the behavior of the flow solutions, we re-derive the scaling relations for combinations of plates of different lengths $L_1$ and $L_2$ ($L_1 \le L_2$) assuming NDPM, similar to the single plate case. Using base units of length as $L_{avg}=1/2(L_1+L_2)$, time as $\sqrt{L_{avg}/g}$, and mass as $\rho_d L_{avg}^3$, we obtain: 
\begin{align*}
F^{D} &= \rho_d g L_{avg}^3 \hat{f}(\mu_s,\mu_f,z/L_{avg},s/L_{avg},w/L_{avg}, L_1/L_2, v/\sqrt{gL_{avg}}).
\end{align*}
With similar assumptions as before in the moderate-depth regime ($F\propto z$),
\begin{align*}
F^{D} &= \rho_d g z L_{avg} w \hat{f}(\mu_s,\mu_f,s/L_{avg},L_1/L_2).
\end{align*}
Defining a non-dimensional force, $\tilde{F} = F^{D}/\rho_d g z L_{avg} w$,  we get    
\begin{align}
\tilde{F} &= \hat{f}(\mu_s,\mu_f,s/L_{avg},L_1/L_2) \label{eq:10}
\end{align}
and for equal plates, $L_1=L_2$, $L_{avg}=L$ and we obtain: 
\begin{align}
\tilde{F} &= \hat{f}(\mu_s,\mu_f,s/L). \label{eq:11}
\end{align}
Therefore, if the NDPM  model suffices to describe the physics of two-plate intrusion, we expect the above relation to describe a master curve that collapases data for plates of various $L$ intruding into the same media.

Based on eq.\ref{eq:10}, for $F\propto z$, we obtain following relations for peak separation ($s_p$) and corresponding peak force ($F_p$) values:
\begin{align}
s_p &= L_{avg}\,\hat{f}_1(L_1/L_2, \mu_s,\mu_f), \ \textrm{and} \\
F_p &= \rho_d g z w L_{avg}\,\hat{f}_2(L_1/L_2, \mu_s,\mu_f) .
\label{eq:sp}
\end{align}
For equal plates ($L_1=L_2=L=L_{avg}$) we obtain
\begin{align}
s_p &= L\,\hat{f}_1(\mu_s, \mu_f), \ \textrm{and} \\
\quad F_p &= \rho_d g z wL\,\hat{f}_2(\mu_s,\mu_f) .
\label{eq:fp}
\end{align}
Note that $F_p$ has units of force per out-of-plane length. Also note that, under these relations, the only material properties that influence the peak force and separation are the friction coefficient(s) and density; the sole length scale comes from the plate itself. This makes physical sense when the smallest in-plane feature (min($w, L,s$)) is sufficiently larger than grain diameter.  However, the exact value of O(1) needs to be calculated experimentally. Swapnil et al.\cite{pravin2020effect} explored the dependence of force-peak separation $s_{p}/d$ with intruder size, $L/d$, when the intruder size is close to the grain diameter, $d$ ($L/d$ range 1--6). Interestingly, their data agree with our proposed linear dependence $s_p\propto L$ at intruder sizes as low as 3-4 grain diameters. \par

\begin{figure}[h!]
\centering
 \includegraphics[trim = 0mm 0mm 0mm 0mm, clip, width=1.0 \linewidth]{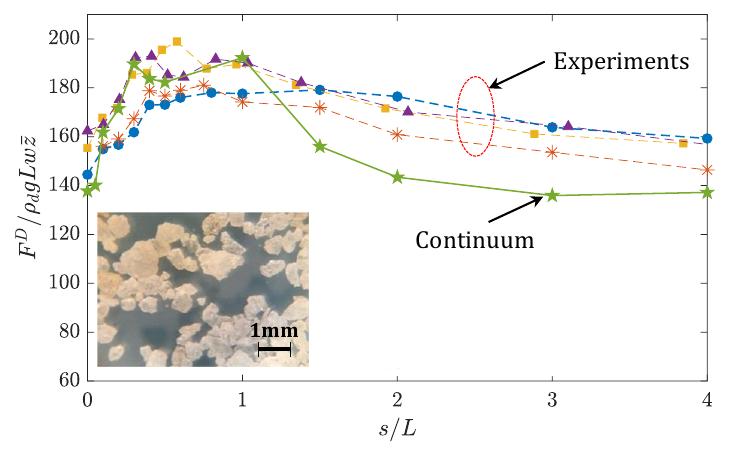}
 \caption{\emph{Case 2: Experimental verification of peak force phenomenon in two-plate intrusions:}. The comparision of experimental data (dotted lines) and calibrated continuum simulations (solid line) for two (equal) plate intrusion experiments. The experiments use plates of width $(L)$ $15\,\textrm{mm}$ (blue data with $\bullet$ marker), $20\,\textrm{mm}$ (orange data with $\ast$ marker), $26\,\textrm{mm}$ (yellow data with  $\blacksquare$ marker), $29\,\textrm{mm}$ (violet  data with  $\blacktriangle$ marker), and the continumm model uses a plate of width $10\,\textrm{mm}$ (green data with $\star$ marker). Quikrete Pool Filter Sand with grain density $\rho_g=2520kg/m^3$ was used (inset shows microscopic view). The effective critical density $\rho_c$ and angle of repose of the sand was found to be $1512 kg/m^3$ and $36^o \pm 1$ resp. All of the paired plates have a $1:5$ horizontal aspect ratio. The continuum results correspond to plane-strain two plate intrusions with drag forces $F^D$ averaged over a depth range of $0.06-0.08$ m for all the cases and an average depth, $\bar{z} = 0.07$ m, $w=1$ m, $g=9.8 m/s^2$, $\rho_d= 1512 kg/m^3 (=\rho_g*\phi)$ is used for scaling the vertical axis. The material properties are calibrated based on experimental data with $\rho_g$=2520 kg/m$^3$, $\phi=0.6$, $\mu_s=0.72$ (=$\tan^{-1}{\theta_{repose}}$).}
 \label{fig:6}
\end{figure}

We also verify these drag force variations and the peak separation scaling relation with new vertical intrusion experiments and compare them to calibrated continuum simulations (see figure \ref{fig:6} for the details). A DENSO VS087 robot arm intruded an apparatus that held pairs of steel plates at various separations into a bed of loosely consolidated Quikrete Pool Filter Sand with grain density $\rho_g=2520$ kg/m$^3$, and effective close-packed density $\rho_d=1512$ kg/m$^3$. The internal friction value of this medium was $\mu_s = 0.72$ based on the angle of repose measurements from experimental tilting tests, with the media globally fluidized for 15 seconds to a loose initial packing fraction of $\phi \approx 0.58$ for all trials. All intrusions were performed at $11$ mm/s, where speed dependence of the force response is negligible. Over 128 trials, the net resistive forces on the pair of intruding plates were measured using an ATI Mini40 force transducer. We observed a satisfactory match between the experiments and the simulations. The fact that the experimental data for different values of $L$ collapse onto a single dimensionless master curve supports the robustness of the scaling relation implied by NDPM in eq \ref{eq:11}. Addionally,{ we observe that although both \ref{fig:5}(A) and Figure \ref{fig:6} show peaks in normalised force responses of the media at low separations, the shapes of the graphs are not 'identical'. This variation is expected because eq \ref{eq:11} indicates that the graph between normalized drag ($\tilde{F}$) and normalized separation ($s/L$) depends on material friction properties ($\mu_s$ and $\mu_f$) and the two cases use different internal friction ($\mu_s$) values.}

We also briefly explore, in our simulations, the effect of changing plate width ratios on the peak separation distance ($s_p$), and the peak separation force ($F_p$) in our study (see figure \ref{fig:5}D). We find that $F_p$ monotonically decreases from a maximum value to $F_{\infty}$ as the plate ratio decreases from $1$ to $0$ (note that plate-ratio $L_1/L_2$ is always $\leq 1$ as $L_1 \leq L_2$ by definition). The normalized peak separation distance ($s_p/L_{avg}$) increases with decreasing plate-ratios ($L_1/L_2$). The peak separation relations are also expected to be a function of $\mu_s$ from scaling analysis (eq. \ref{eq:sp}). Thus, an in-depth shape and material property dependence characterization of this phenomenon is relegated to future study. 
\\


\begin{figure*}[ht!]
\centering
 \includegraphics[trim = 0mm 15mm 0mm 0mm, clip, width=1.0 \linewidth]{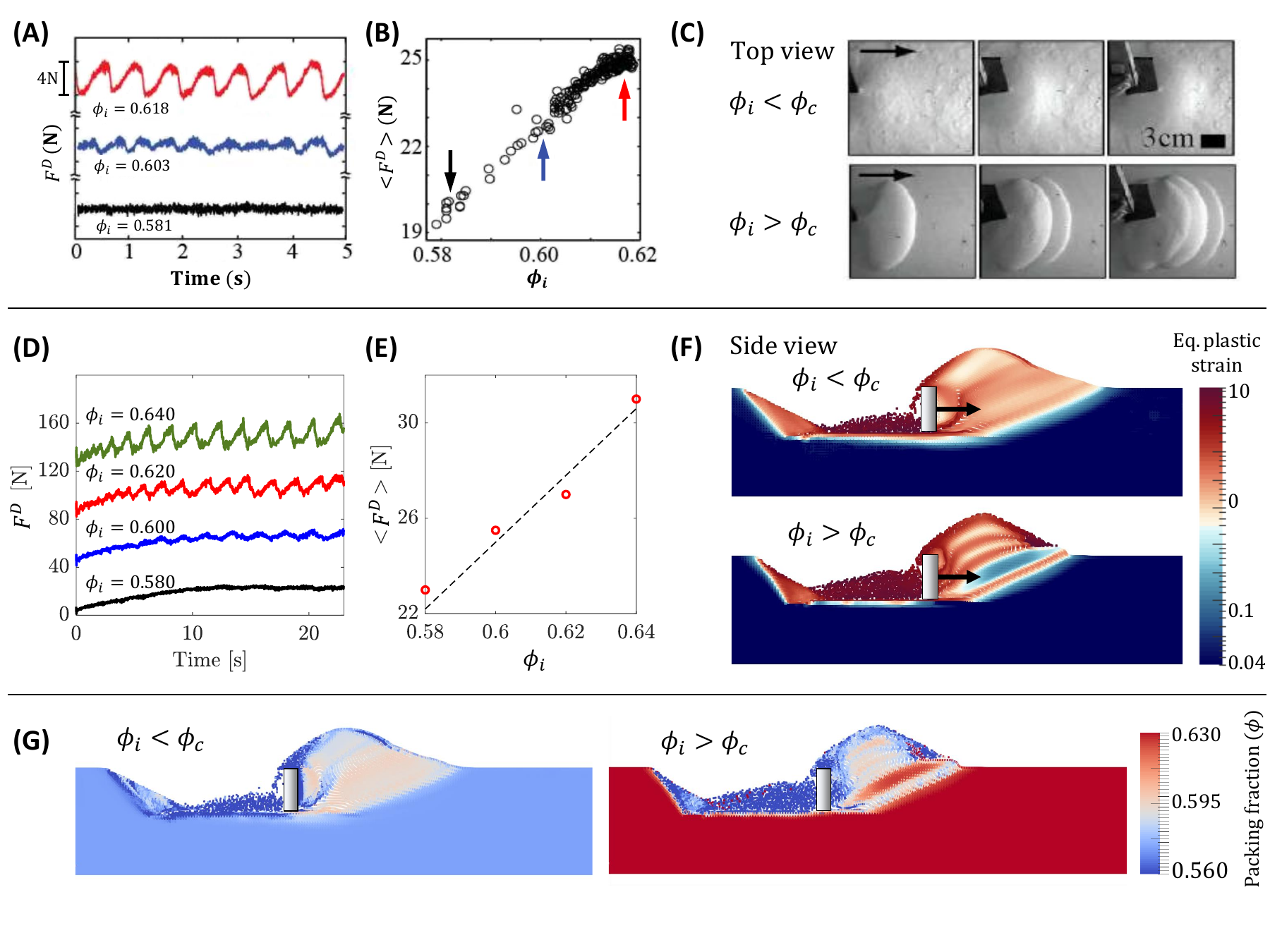}
 \caption{\emph{Case 3: Force fluctuations during plowing {(using DPM model)}}: Variation of drag forces at various initial packing fractions in Gravish et al's \cite{gravish2010force} experiments: \emph{(A)} time variation of drag forces, and \emph{(B)} average drag forces. The average values of forces for three initial packing fractions in \emph{(A)} are shown with corresponding colored arrows in \emph{(B)}. \emph{Corresponding continuum simulations:} \emph{(D)} Time variation of drag forces (force plots for consecutive $\phi_i$ are shifted vertically by $0,$ $40,$ $80,$ and $120$ N respectively for improved visualization), and \emph{(E)} averaged drag forces.  Visualisation of the free surface in under/over compacted granular media: top view from Gravish et al's \cite{gravish2010force} experiments \emph{(C)}, and side view in continuum simulations \emph{(F)}. {\emph{(G)} Variation of material packing fraction from continuum modeling simulations in initially under-compacted ($\phi_i=0.57$) and over-compacted ($\phi_i=0.63$) media cases considered in \emph{(F)}.} The simulations are plane-strain and glass bead properties are used for simulating granular media ($\rho_g$=2520 kg/m$^3$, $\phi_=0.6$, $\mu_c=0.4$, $\chi=5.0$, $\phi_{min}=0.45$). See Movie S3 for visualizing time evolution of material front during plowing of under- and over-compacted media from experiments and simulations. Figure \emph{(A)}, \emph{(B)} and \emph{(C)} are modified from Gravish et al \cite{gravish2010force}. [Photo credits: (C) N. Gravish,  P. B. Umbanhowar and D. I. Goldman, Georgia Institute of Technology]}
 \label{fig:7}
\end{figure*}

\subsection{Drag variations in the plowing of granular media} 
This case takes inspiration from the work of Gravish et al.\cite{gravish2010force} who studied drag force fluctuations in the plowing of granular beds at different initial packing fractions. A Discrete Element Method (DEM) based study of the case was also performed by Kobayakawa et al.\cite{kobayakawa2020interaction}. Both of these studies observed increasing drag force fluctuations with an increasing initial packing fraction of the granular beds. Similar force fluctuations were observed by Kashizadeh and Hambleton \cite{kashizadeh2015numerical} on reduced-order modeling of the plowing processes in sands and by Jin et al.\cite{jin2020small} in the development of a new single-gravity (1-g) small scale testing methodology. As this phenomenon directly relates to the changing density of the media, we use the more detailed DPM model for this case. \par

A schematic representation of this case is given in Figure \ref{fig:1}C. Both of the reported studies were performed in 3D while our simulations are 2D plane-strain. Characterizing the effects of this difference in our studies is difficult, so we do not attempt an exact match. Nevertheless, we do expect the 2D simulations to capture the phenomenon qualitatively and the drag forces to be similar in their magnitudes. We once again use the glass bead material properties used in the previous case. We use a grain density $\rho_g$ of 2520 $\textrm{kg}/\textrm{m}^3$, a critical packing fraction $\phi_c$ of 0.60, a steady-state critical internal friction $\mu_c$ of 0.4, and a scaling coefficient $\chi$ of $2.5$. Similar to previous cases, the plates are modeled as elastic bodies with high elastic modulus to act as rigid bodies. The media/plate interface friction ($\mu_f$) was set to $0.35$.
{We use a $2.4$ m $\times$ $0.4$ m granular bed, and a $4\times10^{-3}$ m spatial resolution ($\Delta x$) for simulating these cases. The plowing plate dimensions are $0.03\times 0.08$ m$^2$}.

Figure \ref{fig:7} shows the variation of horizontal drag forces from the Gravish et al.\cite{gravish2010force} experiments alongside our continuum simulations. The continuum results are scaled proportionally to the out-of-plane width in  Gravish et al.\cite{gravish2010force}. 
The mean drag force and force fluctuations from continuum modeling are plotted in Figure \ref{fig:7}D and E, showing smooth forces transition to larger, fluctuating forces as the initial packing fraction rises above $\phi_c$. The same trends can be seen in experiments, c.f.  Figure \ref{fig:7}A and B. A visual free surface comparison between experiments and the simulations also shows the similarity of flows as $\phi_i$ varies, c.f. Figure \ref{fig:7}C and F. {In the over-compacted case, the continuum results show a stepped pattern forming on the surface but it is not as persistent as the wavy patterns observed in the experiments. This is due to the 2D nature of our simulations. The 2D nature restricts the material from flowing in an out-of-plane direction, which causes the material to flow over the existing waves and overrun the wavy patterns. Such patterns are are otherwise not overrun in 3D experiments and are more visible in over-compacted cases.}

\par 
\begin{figure*}[ht!]
\centering
 \includegraphics[trim =-5mm 67mm 115mm 0mm, clip, width=1.0 \linewidth]{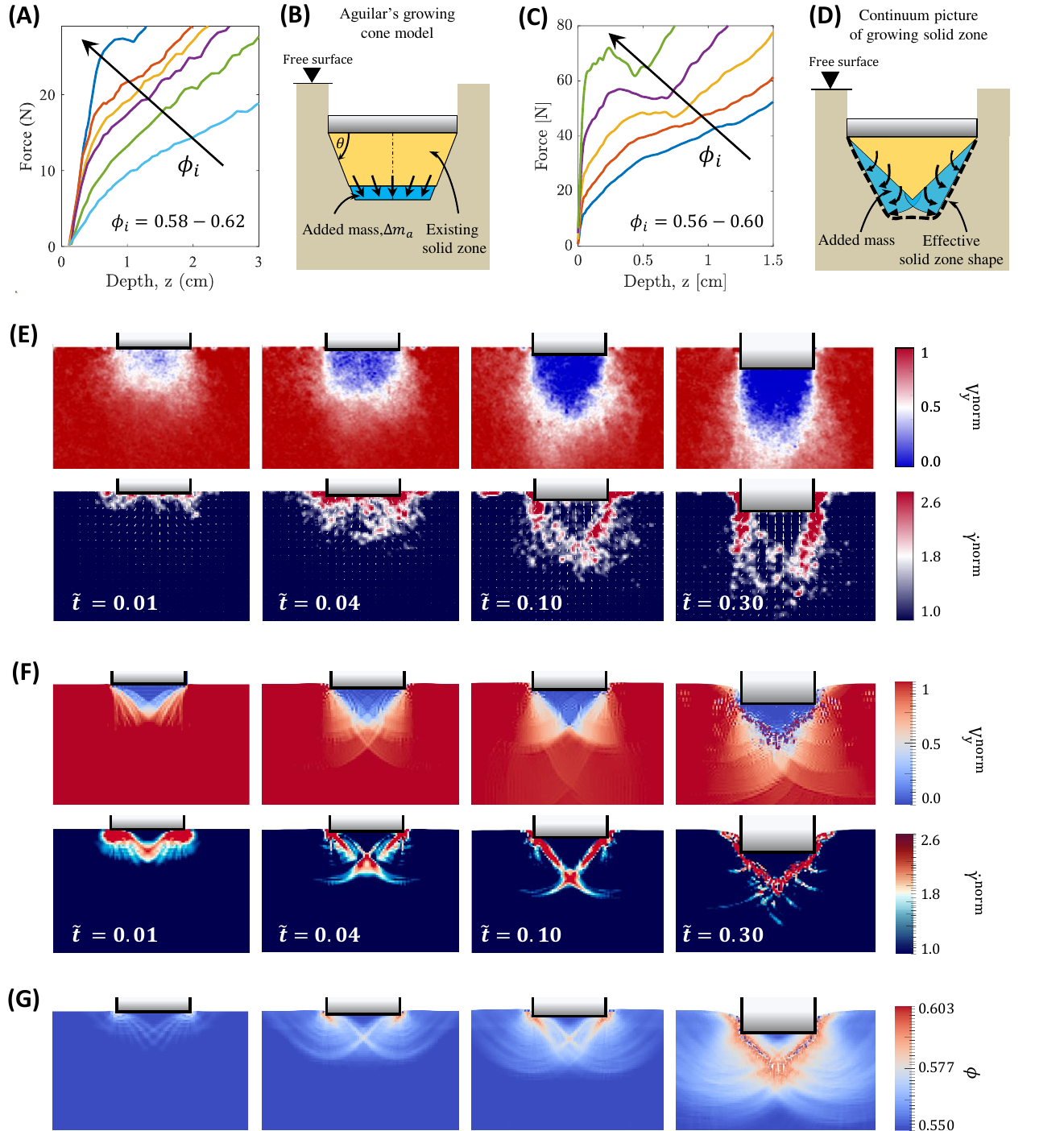}
\caption{\textcolor{black}{\emph{Case 4: Material front development during vertical intrusion  {(using DPM model)}:} \emph{(A)} Experimental data from Aguilar and Goldman \cite{aguilar2016robophysical}. The experiments intruded a circular plate of $0.051$ m diameter into poppy seeds ($\rho_g=1100$ kg/m$^3$) at various initial packing fractions ($\phi_i$). \emph{(B)} Schematic of Aguilar and Goldman's cone growth model\cite{aguilar2016robophysical}. \emph{(C)} \emph{Continuum modelling results:}  Force data from 2D plane strain simulations. We intrude $0.04$ m wide flat plates into material with properties $\rho_g=1100$ kg/m$^3$, $\mu_c=0.53$, $\phi_c=0.60$, $\chi=5.0$, and $\rho_{min}=0.45$ as calibrated to poppy seeds. \emph{(D)} Mechanism of solid zone development apparent from continuum modeling. \emph{(E)} PIV analysis of constant speed plate intrusions. new set of experiments were conducted by intruding $L=40$ mm wide rectangular flat plates next to a clear plexiglass wall at a constant low speed, $v=150\textrm{mm/s}$, in loosely packed poppy seeds. The initial three frames represent the cone growth phase and the last frame ($\tilde{t}\sim0.30$) shows the stability of the cone at later stages of the intrusion. Time is non-dimensionalized ($\tilde{t}=t/t_o$) by $t_o\equiv L/v$. The velocity is normalized by intrusion speed, $v$, and strain rates by $\dot{\gamma}_o \equiv 1/t_0$. We plot these results in the intruder's frame of reference. \emph{(F)} \emph{MPM simulation results:} Vertical velocity and equivalent plastic strain rate fields during vertical intrusion in initially under-compacted media ($\phi_{i} = 0.55$) {for material properties discussed in \emph{(C)}}. We also show the evolution of packing fraction ($\phi$) from continuum modeling in \emph{(G)}. See Movie S4 for visualizing material flow over time for various initial packing fractions ($\phi_i$) including the one considered in \emph{(F)}.}}
 \label{fig:8}
\end{figure*}

The observed shear patterns, free-surface profiles, and force fluctuations shown in Figure \ref{fig:7}A-C are in accord with the shear-softening (shear-strengthening) of over-compacted (under-compacted) granular media that is built-in to the DPM model. In a sample initially under-compacted  ($\phi_i < \phi_c$), shear deformations cause compaction, resulting in higher densities along the sheared regions than in the bulk (see eq.\ref{eq:5}). This density increase results in higher shear strength along the shear zone (see eq.\ref{eq:4}). Thus, further loading in such systems induces shear to occur in the weaker material adjacent to the shear band, which effectively spreads the shearing in such systems. On the other hand, in the over-compacted case, shear deformations dilate the material, resulting in lower density in the sheared region than in the bulk, which results in lower shear strength there.  Hence, continued loading causes shear to accumulate along the thin zone of initial failure causing the appearance of a strong shear band. This process continues until the total force requirement for shearing along the existing band exceeds that for creating a new shear band in the bulk (after which the same process repeats itself).  Thus, in over-compacted media, a visually separable shear band formation pattern occurs (Figure \ref{fig:7}F). In initially over-compacted cases ($\phi_i > \phi_c$), increasing the initial packing fraction of the media results in an increased plate motion requirement between successive shear band formations (due to increased strength of the media in the bulk) and thus the force fluctuation magnitudes increase (and their spatial frequency decreases) with increasing packing fraction (observed also by Gravish et al.\cite{gravish2010force}). {We also plot the variation of the packing fraction in the media from continuum modeling in figure \ref{fig:7}G. The figure provides a visualization of changing packing fraction ahead of the plate in accordance with the smooth versus banded  mechanism explained above.} 
\par

\subsection{Developement of a shear deformation zone in plate intrusions}
This last case takes inspiration from the work of Aguilar and Goldman \cite{aguilar2016robophysical} and highlights the capability of the basic DPM model in capturing the \emph{development} of the flow profile during vertical intrusion of single plates. A schematic representation of the case is given in Figure \ref{fig:1}D. Aguilar and Goldman \cite{aguilar2016robophysical} postulated that the formation of a rigid No Shear Zone (NSZ) ahead of a flat plate during vertical intrusion is an incremental growth process.  In three dimensions, this growth takes the form of a rigid frustum shape which grows from a low height frustum with a fixed base (matching the shape of the intruder) to a fully developed cone/pyramid at the completion of the mechanism. For a circular base, the final shape is a cone. In two dimensions, like our case, this would translate to successive isosceles trapezoids (with larger parallel edges matching the intruder's intruding edge) leading to a wedge shape with the base as the intruder's leading edge. 
 
Figure \ref{fig:8} shows the variation of intrusion forces and successive velocity profiles in under-compacted granular intrusion experiments compared to our continuum simulations. We model the granular media, poppy seeds, with the DPM to incorporate the effect of density transitions. We use material properties for poppy seeds with a grain density $\rho_g$=1100 $\textrm{kg}/\textrm{m}^3$, a critical packing fraction $\phi_c$=0.60, a steady-state critical internal friction $\mu_c$=0.53, and a scaling coefficient $\chi$=5.0. The media/plate interface friction ($\mu_f$) was set to $0.35$. We use a $0.5$ m $\times$ $0.2$ m granular bed, and a $5\times10^{-4}$ m spatial resolution ($\Delta x$) for simulating these cases. The intruding plate dimensions are $0.04$ m $\times$ $0.02$ m. \par
 
 The force trends from simulation qualitatively match the trends from Aguilar's experiments (compare Figure \ref{fig:8}A to \ref{fig:8}C), keeping in mind that Aguilar uses a 3D circular plate while our simulations are in 2D plane-strain. In a set of separate experiments using a rectangular plate intruder and PIV (Figure \ref{fig:8}E) we observed the flow zone development appears similar to continuum results (Figure \ref{fig:8}F).  These trends also agree with general observations from a 3D DEM study done by Feng et al.\cite{feng2019support}. The Aguilar and Goldman \cite{aguilar2016robophysical} study presents a model whereby the rigid cone emerges from a growing rigid frustum of constant base angle that gets progressively longer until converging to the final cone shape (Figure \ref{fig:8}B). However, in our simulations, the zone actually starts as a `short' wedge coincident with the plate bottom, having a large apex angle (expected by a slip-line theory to be approximately $\pi/2 - \tan^{-1}\mu_i$ for $\mu_i$ the friction at the initial density) due to the low initial packing fraction of the media. The rigid front then grows by `fanning out' from the diagonal edges (see Figure \ref{fig:8}D), as the edges represent the zone experiencing maximum shear-compaction and hence the most strengthening. This growth can be observed as well from density variations, shown in Figure \ref{fig:8}G. The growing density of the region results in an increasing but variable internal friction value in the zone below the intruder and results in the development of a progressively sharpening, quasi-rigid trapezoid-like shape under the plate. As the intruder moves deeper, the process converges to a final wedge shape (with a sharper half-angle equal to `$\pi/2 - \tan^{-1}\mu_c$'). Thus the DPM model provides an apt description of the observed behavior in under compacted granular intrusion.  \par

\section{Approach limitations and their implications} 
{While the continuum modeling and the numerical implementation we use in this study are able to represent the considered cases to a sufficient degree, both the model and the method have their limitations. Clearly, as emphasized in the introduction, the phenomena incorporated in a constitutive model limits the behavior the model can capture, and the constitutive relations we use here intentionally exclude certain effects for the benefit of simplicity. Similarly, MPM has known accuracy limitations given by the choice of the grid resolution, material point density, shape functions, and the means of representing contact between domains. For instance, MPM inherently captures a volume average material response everywhere. If during the flow an element has a low number of interior material points, the accuracy of the integration is also diminished. 
But these issues can be overcome with an appropriate choice of shape functions and refined discretization. 
Specifically for these issues, use of more advanced methods such as the hybrid DEM-MPM approach (such as Yue et al. \cite{yue2018hybrid} and Chen et al. \cite{chen2021hybrid}) or dynamic particle enrichment (such as Zhu et al.\cite{zhu2017dynamically}) could be used at the expanse of computation time. The use of smoother and wider shape functions could also help decrease numerical fluctuations often observed in MPM\cite{steffen2008analysis}. }

\section{Conclusions}
{In this work we demonstrated the efficacy of continuum modeling in four cases using two continuum descriptions of granular media, of forced granular intrusion --- (1) depth-dependent force response in horizontal submerged intruder motion; (2) separation-dependent drag variation in parallel plate vertical in-trusion;  (3)  initial density-dependent drag fluctuations in free-surface plowing; and (4) flow zone development in vertical plate intrusions in under compacted granular media (see Figure \ref{fig:1}).} The study shows that relatively simple, friction-based plasticity models capture a large variety of granular intrusion phenomena. Moreover, the models provide a useful macroscopic understanding of granular intrusion processes, which are often primary interests in engineering applications, and remove the additional complexity of trying to determine large-scale physics from grain-scale observations. The simplicity of these continuum models also streamlines this understanding, both by exclusion -- i.e. if such a model works, it implies that mechanisms or effects lying outside the model's formulation are not crucial to the outcome -- and by admitting simple scaling analyses as we have utilized throughout. Such simplifications will certainly limit the accuracy of these models in a variety of cases, but an incremental approach of adding physical augmentations (such as micro-inertial effects, particle size effects, or evolving fabric variables) provides a systematic approach for exploring the underlying physics in diverse cases.  For instance, we do not use $\mu(I)$ rheology in either of the models in this study; the fact that our modeling still captures the observed behaviors indicates micro-inertial effects are not a key mechanism in the observed behaviors. 
We also emphasize that the continuum approach is useful for determining further-reduced models of intrusion, due to the simplicity of its large-scale characterization of systems. One such work on this front was done in Agarwal et al.\cite{agarwal2021surprising}, which develops a generalized, rate-dependent, Dynamic Resistive Force Theory (DRFT) for rapid intrusion of generic intruders. In the future, the work could be extended to three dimensions to extensively compare the computational advantage of using such methods. Furthermore, the continuum treatment could help reconcile granular behavior with similar behaviors observed in other, more standard continua.  
For example, Minetti et al.\cite{minetti2009optimum} reported that during swimming in water, slight separation between fingers increases propulsive thrust, similar to our observation for slightly separated granular intruders \cite{minetti2009optimum,sidelnik2006optimising}.  Comparing and analyzing continuum forms could provide insights into the rationale behind such similarities, as was done in other flow resistance studies\cite{askari2016intrusion}. 

\section*{Conflicts of interest}
There are no conflicts to declare.

\section*{Acknowledgements}
SA and KK acknowledge support from Army Research Office (ARO) grants W911NF1510196, W911NF1810118, and W911NF1910431. AK and DG acknowledge support from
ARO grant {W911NF-18-1-0120
}. We also acknowledge support from the U.S. Army's Tank Automotive Research, Development, and Engineering Center (TARDEC) and thank Dr. Christian Hubicki for his support in conducting vertical intrusion experiments.
%


\balance


\bibliography{rsc} 

\providecommand*{\mcitethebibliography}{\thebibliography}
\csname @ifundefined\endcsname{endmcitethebibliography}
{\let\endmcitethebibliography\endthebibliography}{}
\begin{mcitethebibliography}{72}
\providecommand*{\natexlab}[1]{#1}
\providecommand*{\mciteSetBstSublistMode}[1]{}
\providecommand*{\mciteSetBstMaxWidthForm}[2]{}
\providecommand*{\mciteBstWouldAddEndPuncttrue}
  {\def\EndOfBibitem{\unskip.}}
\providecommand*{\mciteBstWouldAddEndPunctfalse}
  {\let\EndOfBibitem\relax}
\providecommand*{\mciteSetBstMidEndSepPunct}[3]{}
\providecommand*{\mciteSetBstSublistLabelBeginEnd}[3]{}
\providecommand*{\EndOfBibitem}{}
\mciteSetBstSublistMode{f}
\mciteSetBstMaxWidthForm{subitem}
{(\emph{\alph{mcitesubitemcount}})}
\mciteSetBstSublistLabelBeginEnd{\mcitemaxwidthsubitemform\space}
{\relax}{\relax}

\bibitem[Soliman \emph{et~al.}(1976)Soliman, Reid, and
  Johnson]{soliman1976effect}
A.~Soliman, S.~Reid and W.~Johnson, \emph{International Journal of Mechanical
  Sciences}, 1976, \textbf{18}, 279--284\relax
\mciteBstWouldAddEndPuncttrue
\mciteSetBstMidEndSepPunct{\mcitedefaultmidpunct}
{\mcitedefaultendpunct}{\mcitedefaultseppunct}\relax
\EndOfBibitem
\bibitem[Omidvar \emph{et~al.}(2014)Omidvar, Iskander, and
  Bless]{omidvar2014response}
M.~Omidvar, M.~Iskander and S.~Bless, \emph{International Journal of Impact
  Engineering}, 2014, \textbf{66}, 60--82\relax
\mciteBstWouldAddEndPuncttrue
\mciteSetBstMidEndSepPunct{\mcitedefaultmidpunct}
{\mcitedefaultendpunct}{\mcitedefaultseppunct}\relax
\EndOfBibitem
\bibitem[Ortiz \emph{et~al.}(2019)Ortiz, Gravish, and Tolley]{ortiz2019soft}
D.~Ortiz, N.~Gravish and M.~T. Tolley, \emph{IEEE Robotics and Automation
  Letters}, 2019, \textbf{4}, 2630--2636\relax
\mciteBstWouldAddEndPuncttrue
\mciteSetBstMidEndSepPunct{\mcitedefaultmidpunct}
{\mcitedefaultendpunct}{\mcitedefaultseppunct}\relax
\EndOfBibitem
\bibitem[Van Der~Meer(2017)]{van2017impact}
D.~Van Der~Meer, \emph{Annual review of fluid mechanics}, 2017, \textbf{49},
  463\relax
\mciteBstWouldAddEndPuncttrue
\mciteSetBstMidEndSepPunct{\mcitedefaultmidpunct}
{\mcitedefaultendpunct}{\mcitedefaultseppunct}\relax
\EndOfBibitem
\bibitem[Andreotti \emph{et~al.}(2013)Andreotti, Forterre, and
  Pouliquen]{andreotti2013granular}
B.~Andreotti, Y.~Forterre and O.~Pouliquen, \emph{Granular media: between fluid
  and solid}, Cambridge University Press, 2013\relax
\mciteBstWouldAddEndPuncttrue
\mciteSetBstMidEndSepPunct{\mcitedefaultmidpunct}
{\mcitedefaultendpunct}{\mcitedefaultseppunct}\relax
\EndOfBibitem
\bibitem[Wood(1990)]{wood1990soil}
D.~M. Wood, \emph{Soil behaviour and critical state soil mechanics}, Cambridge
  University Press, 1990\relax
\mciteBstWouldAddEndPuncttrue
\mciteSetBstMidEndSepPunct{\mcitedefaultmidpunct}
{\mcitedefaultendpunct}{\mcitedefaultseppunct}\relax
\EndOfBibitem
\bibitem[Jensen \emph{et~al.}(1999)Jensen, Bosscher, Plesha, and
  Edil]{jensen1999simulation}
R.~P. Jensen, P.~J. Bosscher, M.~E. Plesha and T.~B. Edil, \emph{International
  Journal for Numerical and Analytical Methods in Geomechanics}, 1999,
  \textbf{23}, 531--547\relax
\mciteBstWouldAddEndPuncttrue
\mciteSetBstMidEndSepPunct{\mcitedefaultmidpunct}
{\mcitedefaultendpunct}{\mcitedefaultseppunct}\relax
\EndOfBibitem
\bibitem[Jensen \emph{et~al.}(2001)Jensen, Plesha, Edil, Bosscher, and
  Kahla]{jensen2001simulation}
R.~P. Jensen, M.~E. Plesha, T.~B. Edil, P.~J. Bosscher and N.~B. Kahla,
  \emph{International Journal of Geomechanics}, 2001, \textbf{1}, 21--39\relax
\mciteBstWouldAddEndPuncttrue
\mciteSetBstMidEndSepPunct{\mcitedefaultmidpunct}
{\mcitedefaultendpunct}{\mcitedefaultseppunct}\relax
\EndOfBibitem
\bibitem[Brilliantov \emph{et~al.}(1996)Brilliantov, Spahn, Hertzsch, and
  P{\"o}schel]{brilliantov1996model}
N.~V. Brilliantov, F.~Spahn, J.-M. Hertzsch and T.~P{\"o}schel, \emph{Physical
  review E}, 1996, \textbf{53}, 5382\relax
\mciteBstWouldAddEndPuncttrue
\mciteSetBstMidEndSepPunct{\mcitedefaultmidpunct}
{\mcitedefaultendpunct}{\mcitedefaultseppunct}\relax
\EndOfBibitem
\bibitem[Silbert \emph{et~al.}(2001)Silbert, Erta{\c{s}}, Grest, Halsey,
  Levine, and Plimpton]{silbert2001granular}
L.~E. Silbert, D.~Erta{\c{s}}, G.~S. Grest, T.~C. Halsey, D.~Levine and S.~J.
  Plimpton, \emph{Physical Review E}, 2001, \textbf{64}, 051302\relax
\mciteBstWouldAddEndPuncttrue
\mciteSetBstMidEndSepPunct{\mcitedefaultmidpunct}
{\mcitedefaultendpunct}{\mcitedefaultseppunct}\relax
\EndOfBibitem
\bibitem[Kamrin and Koval(2012)]{kamrin2012nonlocal}
K.~Kamrin and G.~Koval, \emph{Physical Review Letters}, 2012, \textbf{108},
  178301\relax
\mciteBstWouldAddEndPuncttrue
\mciteSetBstMidEndSepPunct{\mcitedefaultmidpunct}
{\mcitedefaultendpunct}{\mcitedefaultseppunct}\relax
\EndOfBibitem
\bibitem[Kim and Kamrin(2020)]{kim2020power}
S.~Kim and K.~Kamrin, \emph{Physical Review Letters}, 2020, \textbf{125 (8)},
  088002\relax
\mciteBstWouldAddEndPuncttrue
\mciteSetBstMidEndSepPunct{\mcitedefaultmidpunct}
{\mcitedefaultendpunct}{\mcitedefaultseppunct}\relax
\EndOfBibitem
\bibitem[Jajcevic \emph{et~al.}(2013)Jajcevic, Siegmann, Radeke, and
  Khinast]{jajcevic2013large}
D.~Jajcevic, E.~Siegmann, C.~Radeke and J.~G. Khinast, \emph{Chemical
  Engineering Science}, 2013, \textbf{98}, 298--310\relax
\mciteBstWouldAddEndPuncttrue
\mciteSetBstMidEndSepPunct{\mcitedefaultmidpunct}
{\mcitedefaultendpunct}{\mcitedefaultseppunct}\relax
\EndOfBibitem
\bibitem[Zhong \emph{et~al.}(2016)Zhong, Yu, Liu, Tong, and
  Zhang]{zhong2016cfd}
W.~Zhong, A.~Yu, X.~Liu, Z.~Tong and H.~Zhang, \emph{Powder Technology}, 2016,
  \textbf{302}, 108--152\relax
\mciteBstWouldAddEndPuncttrue
\mciteSetBstMidEndSepPunct{\mcitedefaultmidpunct}
{\mcitedefaultendpunct}{\mcitedefaultseppunct}\relax
\EndOfBibitem
\bibitem[Kelly \emph{et~al.}(2020)Kelly, Olsen, and Negrut]{kelly2019billion}
C.~Kelly, N.~Olsen and D.~Negrut, \emph{Multibody Systems Dynamics}, 2020,
  \textbf{5}, 355--379\relax
\mciteBstWouldAddEndPuncttrue
\mciteSetBstMidEndSepPunct{\mcitedefaultmidpunct}
{\mcitedefaultendpunct}{\mcitedefaultseppunct}\relax
\EndOfBibitem
\bibitem[Vriend \emph{et~al.}(2013)Vriend, McElwaine, Sovilla, Keylock, Ash,
  and Brennan]{vriend2013high}
N.~Vriend, J.~McElwaine, B.~Sovilla, C.~Keylock, M.~Ash and P.~Brennan,
  \emph{Geophysical Research Letters}, 2013, \textbf{40}, 727--731\relax
\mciteBstWouldAddEndPuncttrue
\mciteSetBstMidEndSepPunct{\mcitedefaultmidpunct}
{\mcitedefaultendpunct}{\mcitedefaultseppunct}\relax
\EndOfBibitem
\bibitem[Daniels and Hayman(2008)]{daniels2008force}
K.~E. Daniels and N.~W. Hayman, \emph{Journal of Geophysical Research: Solid
  Earth}, 2008, \textbf{113}, B11\relax
\mciteBstWouldAddEndPuncttrue
\mciteSetBstMidEndSepPunct{\mcitedefaultmidpunct}
{\mcitedefaultendpunct}{\mcitedefaultseppunct}\relax
\EndOfBibitem
\bibitem[Agarwal \emph{et~al.}(2019)Agarwal, Senatore, Zhang, Kingsbury,
  Iagnemma, Goldman, and Kamrin]{agarwal2019modeling}
S.~Agarwal, C.~Senatore, T.~Zhang, M.~Kingsbury, K.~Iagnemma, D.~I. Goldman and
  K.~Kamrin, \emph{Journal of Terramechanics}, 2019, \textbf{85}, 1--14\relax
\mciteBstWouldAddEndPuncttrue
\mciteSetBstMidEndSepPunct{\mcitedefaultmidpunct}
{\mcitedefaultendpunct}{\mcitedefaultseppunct}\relax
\EndOfBibitem
\bibitem[Li \emph{et~al.}(2013)Li, Zhang, and Goldman]{li2013terradynamics}
C.~Li, T.~Zhang and D.~I. Goldman, \emph{science}, 2013, \textbf{339},
  1408--1412\relax
\mciteBstWouldAddEndPuncttrue
\mciteSetBstMidEndSepPunct{\mcitedefaultmidpunct}
{\mcitedefaultendpunct}{\mcitedefaultseppunct}\relax
\EndOfBibitem
\bibitem[Agarwal \emph{et~al.}(2021)Agarwal, Karsai, Goldman, and
  Kamrin]{agarwal2021surprising}
S.~Agarwal, A.~Karsai, D.~I. Goldman and K.~Kamrin, \emph{Science Advances},
  2021, \textbf{7}, eabe0631\relax
\mciteBstWouldAddEndPuncttrue
\mciteSetBstMidEndSepPunct{\mcitedefaultmidpunct}
{\mcitedefaultendpunct}{\mcitedefaultseppunct}\relax
\EndOfBibitem
\bibitem[Mahabadi and Jang(2017)]{mahabadi2017impact}
N.~Mahabadi and J.~Jang, \emph{Applied Physics Letters}, 2017, \textbf{110},
  041907\relax
\mciteBstWouldAddEndPuncttrue
\mciteSetBstMidEndSepPunct{\mcitedefaultmidpunct}
{\mcitedefaultendpunct}{\mcitedefaultseppunct}\relax
\EndOfBibitem
\bibitem[Brilliantov and P{\"o}schel(2010)]{brilliantov2010kinetic}
N.~V. Brilliantov and T.~P{\"o}schel, \emph{Kinetic theory of granular gases},
  Oxford University Press, 2010\relax
\mciteBstWouldAddEndPuncttrue
\mciteSetBstMidEndSepPunct{\mcitedefaultmidpunct}
{\mcitedefaultendpunct}{\mcitedefaultseppunct}\relax
\EndOfBibitem
\bibitem[Jenkins and Richman(1985)]{jenkins1985kinetic}
J.~T. Jenkins and M.~Richman, \emph{The Physics of fluids}, 1985, \textbf{28},
  3485--3494\relax
\mciteBstWouldAddEndPuncttrue
\mciteSetBstMidEndSepPunct{\mcitedefaultmidpunct}
{\mcitedefaultendpunct}{\mcitedefaultseppunct}\relax
\EndOfBibitem
\bibitem[Schofield and Wroth(1968)]{schofield1968critical}
A.~Schofield and P.~Wroth, \emph{Critical state soil mechanics}, McGraw-hill,
  1968\relax
\mciteBstWouldAddEndPuncttrue
\mciteSetBstMidEndSepPunct{\mcitedefaultmidpunct}
{\mcitedefaultendpunct}{\mcitedefaultseppunct}\relax
\EndOfBibitem
\bibitem[Yu(2008)]{yu2008non}
H.-S. Yu, Keynote lecture. Proc. 12th Int. Conf. IACMAG, Goa, India, 2008, pp.
  361--378\relax
\mciteBstWouldAddEndPuncttrue
\mciteSetBstMidEndSepPunct{\mcitedefaultmidpunct}
{\mcitedefaultendpunct}{\mcitedefaultseppunct}\relax
\EndOfBibitem
\bibitem[Jop \emph{et~al.}(2006)Jop, Forterre, and
  Pouliquen]{jop2006constitutive}
P.~Jop, Y.~Forterre and O.~Pouliquen, \emph{Nature}, 2006, \textbf{441},
  727--730\relax
\mciteBstWouldAddEndPuncttrue
\mciteSetBstMidEndSepPunct{\mcitedefaultmidpunct}
{\mcitedefaultendpunct}{\mcitedefaultseppunct}\relax
\EndOfBibitem
\bibitem[Wiacek \emph{et~al.}(2011)Wiacek, Molenda,
  Horabik,\emph{et~al.}]{wiacek2011mechanical}
J.~Wiacek, M.~Molenda, J.~Horabik \emph{et~al.}, \emph{Mechanical properties of
  granular agro-materials. Continuum and discrete approach}, Polska Akademia
  Nauk, 2011\relax
\mciteBstWouldAddEndPuncttrue
\mciteSetBstMidEndSepPunct{\mcitedefaultmidpunct}
{\mcitedefaultendpunct}{\mcitedefaultseppunct}\relax
\EndOfBibitem
\bibitem[Henann and Kamrin(2013)]{henann2013predictive}
D.~L. Henann and K.~Kamrin, \emph{Proceedings of the National Academy of
  Sciences}, 2013, \textbf{110}, 6730--6735\relax
\mciteBstWouldAddEndPuncttrue
\mciteSetBstMidEndSepPunct{\mcitedefaultmidpunct}
{\mcitedefaultendpunct}{\mcitedefaultseppunct}\relax
\EndOfBibitem
\bibitem[Kamrin(2019)]{kamrin2019non}
K.~Kamrin, \emph{Frontiers in Physics}, 2019, \textbf{7}, 116\relax
\mciteBstWouldAddEndPuncttrue
\mciteSetBstMidEndSepPunct{\mcitedefaultmidpunct}
{\mcitedefaultendpunct}{\mcitedefaultseppunct}\relax
\EndOfBibitem
\bibitem[Dunatunga and Kamrin(2015)]{dunatunga2015continuum}
S.~Dunatunga and K.~Kamrin, \emph{Journal of Fluid Mechanics}, 2015,
  \textbf{779}, 483--513\relax
\mciteBstWouldAddEndPuncttrue
\mciteSetBstMidEndSepPunct{\mcitedefaultmidpunct}
{\mcitedefaultendpunct}{\mcitedefaultseppunct}\relax
\EndOfBibitem
\bibitem[Daviet and Bertails-Descoubes(2016)]{daviet2016semi}
G.~Daviet and F.~Bertails-Descoubes, \emph{ACM Transactions on Graphics (TOG)},
  2016, \textbf{35}, 1--13\relax
\mciteBstWouldAddEndPuncttrue
\mciteSetBstMidEndSepPunct{\mcitedefaultmidpunct}
{\mcitedefaultendpunct}{\mcitedefaultseppunct}\relax
\EndOfBibitem
\bibitem[Daviet and Bertails-Descoubes(2016)]{daviet2016nonsmooth}
G.~Daviet and F.~Bertails-Descoubes, \emph{Journal of Non-Newtonian Fluid
  Mechanics}, 2016, \textbf{234}, 15--35\relax
\mciteBstWouldAddEndPuncttrue
\mciteSetBstMidEndSepPunct{\mcitedefaultmidpunct}
{\mcitedefaultendpunct}{\mcitedefaultseppunct}\relax
\EndOfBibitem
\bibitem[D{\'a}valos \emph{et~al.}(2015)D{\'a}valos, Cante, Hern{\'a}ndez, and
  Oliver]{davalos2015numerical}
C.~D{\'a}valos, J.~Cante, J.~Hern{\'a}ndez and J.~Oliver, \emph{International
  Journal of Solids and Structures}, 2015, \textbf{71}, 99--125\relax
\mciteBstWouldAddEndPuncttrue
\mciteSetBstMidEndSepPunct{\mcitedefaultmidpunct}
{\mcitedefaultendpunct}{\mcitedefaultseppunct}\relax
\EndOfBibitem
\bibitem[Peng \emph{et~al.}(2016)Peng, Guo, Wu, and Wang]{peng2016unified}
C.~Peng, X.~Guo, W.~Wu and Y.~Wang, \emph{Acta Geotechnica}, 2016, \textbf{11},
  1231--1247\relax
\mciteBstWouldAddEndPuncttrue
\mciteSetBstMidEndSepPunct{\mcitedefaultmidpunct}
{\mcitedefaultendpunct}{\mcitedefaultseppunct}\relax
\EndOfBibitem
\bibitem[Dunatunga and Kamrin(2017)]{dunatunga2017continuum}
S.~Dunatunga and K.~Kamrin, \emph{Journal of the Mechanics and Physics of
  Solids}, 2017, \textbf{100}, 45--60\relax
\mciteBstWouldAddEndPuncttrue
\mciteSetBstMidEndSepPunct{\mcitedefaultmidpunct}
{\mcitedefaultendpunct}{\mcitedefaultseppunct}\relax
\EndOfBibitem
\bibitem[Taylor(1948)]{taylor1948fundamentals}
D.~W. Taylor, \emph{Fundamentals of soil mechanics}, LWW, 1948, vol.~66\relax
\mciteBstWouldAddEndPuncttrue
\mciteSetBstMidEndSepPunct{\mcitedefaultmidpunct}
{\mcitedefaultendpunct}{\mcitedefaultseppunct}\relax
\EndOfBibitem
\bibitem[Agarwal(2020)]{agarwal2020}
S.~Agarwal, \emph{Continuum Modeling of Granular Flows and Intrusions}, 2020,
  MATLAB Central File Exchange\relax
\mciteBstWouldAddEndPuncttrue
\mciteSetBstMidEndSepPunct{\mcitedefaultmidpunct}
{\mcitedefaultendpunct}{\mcitedefaultseppunct}\relax
\EndOfBibitem
\bibitem[Sulsky \emph{et~al.}(1994)Sulsky, Chen, and
  Schreyer]{sulsky1994particle}
D.~Sulsky, Z.~Chen and H.~L. Schreyer, \emph{Computer methods in applied
  mechanics and engineering}, 1994, \textbf{118}, 179--196\relax
\mciteBstWouldAddEndPuncttrue
\mciteSetBstMidEndSepPunct{\mcitedefaultmidpunct}
{\mcitedefaultendpunct}{\mcitedefaultseppunct}\relax
\EndOfBibitem
\bibitem[Jarray \emph{et~al.}(2019)Jarray, Shi, Scheper, Habibi, and
  Luding]{jarray2019cohesion}
A.~Jarray, H.~Shi, B.~J. Scheper, M.~Habibi and S.~Luding, \emph{Scientific
  reports}, 2019, \textbf{9}, 1--12\relax
\mciteBstWouldAddEndPuncttrue
\mciteSetBstMidEndSepPunct{\mcitedefaultmidpunct}
{\mcitedefaultendpunct}{\mcitedefaultseppunct}\relax
\EndOfBibitem
\bibitem[Aguilar \emph{et~al.}(2016)Aguilar, Zhang, Qian, Kingsbury, McInroe,
  Mazouchova, Li, Maladen, Gong, Travers,\emph{et~al.}]{aguilar2016review}
J.~Aguilar, T.~Zhang, F.~Qian, M.~Kingsbury, B.~McInroe, N.~Mazouchova, C.~Li,
  R.~Maladen, C.~Gong, M.~Travers \emph{et~al.}, \emph{Reports on Progress in
  Physics}, 2016, \textbf{79}, 110001\relax
\mciteBstWouldAddEndPuncttrue
\mciteSetBstMidEndSepPunct{\mcitedefaultmidpunct}
{\mcitedefaultendpunct}{\mcitedefaultseppunct}\relax
\EndOfBibitem
\bibitem[Li \emph{et~al.}(2019)Li, Lai, Wang, Yang, Jiang, Li, Wang, and
  Yang]{li2019review}
H.~Li, Y.~Lai, L.~Wang, X.~Yang, N.~Jiang, L.~Li, C.~Wang and B.~Yang,
  \emph{Cold regions science and technology}, 2019, \textbf{157},
  171--186\relax
\mciteBstWouldAddEndPuncttrue
\mciteSetBstMidEndSepPunct{\mcitedefaultmidpunct}
{\mcitedefaultendpunct}{\mcitedefaultseppunct}\relax
\EndOfBibitem
\bibitem[Zhang \emph{et~al.}(2015)Zhang, Sheng, Kouretzis, Krabbenhoft, and
  Sloan]{zhang2015numerical}
X.~Zhang, D.~Sheng, G.~P. Kouretzis, K.~Krabbenhoft and S.~W. Sloan,
  \emph{Physical Review E}, 2015, \textbf{91}, 022204\relax
\mciteBstWouldAddEndPuncttrue
\mciteSetBstMidEndSepPunct{\mcitedefaultmidpunct}
{\mcitedefaultendpunct}{\mcitedefaultseppunct}\relax
\EndOfBibitem
\bibitem[Guillard \emph{et~al.}(2014)Guillard, Forterre, and
  Pouliquen]{guillard2014lift}
F.~Guillard, Y.~Forterre and O.~Pouliquen, \emph{Physics of Fluids}, 2014,
  \textbf{26}, 043301\relax
\mciteBstWouldAddEndPuncttrue
\mciteSetBstMidEndSepPunct{\mcitedefaultmidpunct}
{\mcitedefaultendpunct}{\mcitedefaultseppunct}\relax
\EndOfBibitem
\bibitem[Wu \emph{et~al.}(2020)Wu, Kouretzis, Suwal, Ansari, and
  Sloan]{wu2020shallow}
J.~Wu, G.~Kouretzis, L.~Suwal, Y.~Ansari and S.~W. Sloan, \emph{Canadian
  Geotechnical Journal}, 2020, \textbf{57}, 1472--1483\relax
\mciteBstWouldAddEndPuncttrue
\mciteSetBstMidEndSepPunct{\mcitedefaultmidpunct}
{\mcitedefaultendpunct}{\mcitedefaultseppunct}\relax
\EndOfBibitem
\bibitem[Ding \emph{et~al.}(2011)Ding, Gravish, and Goldman]{ding2011drag}
Y.~Ding, N.~Gravish and D.~I. Goldman, \emph{Physical review letters}, 2011,
  \textbf{106}, 028001\relax
\mciteBstWouldAddEndPuncttrue
\mciteSetBstMidEndSepPunct{\mcitedefaultmidpunct}
{\mcitedefaultendpunct}{\mcitedefaultseppunct}\relax
\EndOfBibitem
\bibitem[Klinkm{\"u}ller \emph{et~al.}(2016)Klinkm{\"u}ller, Schreurs, Rosenau,
  and Kemnitz]{klinkmuller2016properties}
M.~Klinkm{\"u}ller, G.~Schreurs, M.~Rosenau and H.~Kemnitz,
  \emph{Tectonophysics}, 2016, \textbf{684}, 23--38\relax
\mciteBstWouldAddEndPuncttrue
\mciteSetBstMidEndSepPunct{\mcitedefaultmidpunct}
{\mcitedefaultendpunct}{\mcitedefaultseppunct}\relax
\EndOfBibitem
\bibitem[De~La~Cruz and Caballero-Robledo(2016)]{de2016lift}
R.~L. De~La~Cruz and G.~Caballero-Robledo, \emph{Journal of Fluid Mechanics},
  2016, \textbf{800}, 248--263\relax
\mciteBstWouldAddEndPuncttrue
\mciteSetBstMidEndSepPunct{\mcitedefaultmidpunct}
{\mcitedefaultendpunct}{\mcitedefaultseppunct}\relax
\EndOfBibitem
\bibitem[Merceron \emph{et~al.}(2018)Merceron, Sauret, and
  Jop]{merceron2018cooperative}
A.~Merceron, A.~Sauret and P.~Jop, \emph{EPL (Europhysics Letters)}, 2018,
  \textbf{121}, 34005\relax
\mciteBstWouldAddEndPuncttrue
\mciteSetBstMidEndSepPunct{\mcitedefaultmidpunct}
{\mcitedefaultendpunct}{\mcitedefaultseppunct}\relax
\EndOfBibitem
\bibitem[Pravin \emph{et~al.}(2020)Pravin, Chang, Han, London, Goldman, Jaeger,
  and Hsieh]{pravin2020effect}
S.~Pravin, B.~Chang, E.~Han, L.~London, D.~I. Goldman, H.~M. Jaeger and S.~T.
  Hsieh, \emph{arXiv preprint arXiv:2010.15172}, 2020\relax
\mciteBstWouldAddEndPuncttrue
\mciteSetBstMidEndSepPunct{\mcitedefaultmidpunct}
{\mcitedefaultendpunct}{\mcitedefaultseppunct}\relax
\EndOfBibitem
\bibitem[Clark and Behringer(2013)]{clark2013granular}
A.~H. Clark and R.~P. Behringer, \emph{EPL (Europhysics Letters)}, 2013,
  \textbf{101}, 64001\relax
\mciteBstWouldAddEndPuncttrue
\mciteSetBstMidEndSepPunct{\mcitedefaultmidpunct}
{\mcitedefaultendpunct}{\mcitedefaultseppunct}\relax
\EndOfBibitem
\bibitem[Roth \emph{et~al.}(2021)Roth, Han, and Jaeger]{PhysRevLett.126.218001}
L.~K. Roth, E.~Han and H.~M. Jaeger, \emph{Phys. Rev. Lett.}, 2021,
  \textbf{126}, 218001\relax
\mciteBstWouldAddEndPuncttrue
\mciteSetBstMidEndSepPunct{\mcitedefaultmidpunct}
{\mcitedefaultendpunct}{\mcitedefaultseppunct}\relax
\EndOfBibitem
\bibitem[Katsuragi and Durian(2007)]{katsuragi2007unified}
H.~Katsuragi and D.~J. Durian, \emph{Nature physics}, 2007, \textbf{3},
  420--423\relax
\mciteBstWouldAddEndPuncttrue
\mciteSetBstMidEndSepPunct{\mcitedefaultmidpunct}
{\mcitedefaultendpunct}{\mcitedefaultseppunct}\relax
\EndOfBibitem
\bibitem[Brzinski~III \emph{et~al.}(2013)Brzinski~III, Mayor, and
  Durian]{brzinski2013depth}
T.~A. Brzinski~III, P.~Mayor and D.~J. Durian, \emph{Physical review letters},
  2013, \textbf{111}, 168002\relax
\mciteBstWouldAddEndPuncttrue
\mciteSetBstMidEndSepPunct{\mcitedefaultmidpunct}
{\mcitedefaultendpunct}{\mcitedefaultseppunct}\relax
\EndOfBibitem
\bibitem[Kang \emph{et~al.}(2018)Kang, Feng, Liu, and
  Blumenfeld]{kang2018archimedes}
W.~Kang, Y.~Feng, C.~Liu and R.~Blumenfeld, \emph{Nature communications}, 2018,
  \textbf{9}, 1--9\relax
\mciteBstWouldAddEndPuncttrue
\mciteSetBstMidEndSepPunct{\mcitedefaultmidpunct}
{\mcitedefaultendpunct}{\mcitedefaultseppunct}\relax
\EndOfBibitem
\bibitem[Bester and Behringer(2017)]{bester2017collisional}
C.~S. Bester and R.~P. Behringer, \emph{Physical Review E}, 2017, \textbf{95},
  032906\relax
\mciteBstWouldAddEndPuncttrue
\mciteSetBstMidEndSepPunct{\mcitedefaultmidpunct}
{\mcitedefaultendpunct}{\mcitedefaultseppunct}\relax
\EndOfBibitem
\bibitem[Sokolovski{\u\i}(1960)]{sokolovskiui1960statics}
V.~V. Sokolovski{\u\i}, \emph{Statics of soil media}, Butterworths Scientific
  Publications, 1960\relax
\mciteBstWouldAddEndPuncttrue
\mciteSetBstMidEndSepPunct{\mcitedefaultmidpunct}
{\mcitedefaultendpunct}{\mcitedefaultseppunct}\relax
\EndOfBibitem
\bibitem[Bower(2009)]{bower2009applied}
A.~F. Bower, \emph{Chapter 6, Applied mechanics of Solids}, CRC press,
  2009\relax
\mciteBstWouldAddEndPuncttrue
\mciteSetBstMidEndSepPunct{\mcitedefaultmidpunct}
{\mcitedefaultendpunct}{\mcitedefaultseppunct}\relax
\EndOfBibitem
\bibitem[Aguilar and Goldman(2016)]{aguilar2016robophysical}
J.~Aguilar and D.~I. Goldman, \emph{Nature Physics}, 2016, \textbf{12},
  278--283\relax
\mciteBstWouldAddEndPuncttrue
\mciteSetBstMidEndSepPunct{\mcitedefaultmidpunct}
{\mcitedefaultendpunct}{\mcitedefaultseppunct}\relax
\EndOfBibitem
\bibitem[Prandtl(1920)]{prandtl1920concerning}
L.~Prandtl, \emph{Nachr. Ges. Wiss. Gottingen}, 1920,  74--85\relax
\mciteBstWouldAddEndPuncttrue
\mciteSetBstMidEndSepPunct{\mcitedefaultmidpunct}
{\mcitedefaultendpunct}{\mcitedefaultseppunct}\relax
\EndOfBibitem
\bibitem[Hill(1949)]{hill1949plastic}
R.~Hill, \emph{The Quarterly Journal of Mechanics and Applied Mathematics},
  1949, \textbf{2}, 40--52\relax
\mciteBstWouldAddEndPuncttrue
\mciteSetBstMidEndSepPunct{\mcitedefaultmidpunct}
{\mcitedefaultendpunct}{\mcitedefaultseppunct}\relax
\EndOfBibitem
\bibitem[Gravish \emph{et~al.}(2010)Gravish, Umbanhowar, and
  Goldman]{gravish2010force}
N.~Gravish, P.~B. Umbanhowar and D.~I. Goldman, \emph{Physical review letters},
  2010, \textbf{105}, 128301\relax
\mciteBstWouldAddEndPuncttrue
\mciteSetBstMidEndSepPunct{\mcitedefaultmidpunct}
{\mcitedefaultendpunct}{\mcitedefaultseppunct}\relax
\EndOfBibitem
\bibitem[Kobayakawa \emph{et~al.}(2020)Kobayakawa, Miyai, Tsuji, and
  Tanaka]{kobayakawa2020interaction}
M.~Kobayakawa, S.~Miyai, T.~Tsuji and T.~Tanaka, \emph{Journal of
  Terramechanics}, 2020, \textbf{90}, 3--10\relax
\mciteBstWouldAddEndPuncttrue
\mciteSetBstMidEndSepPunct{\mcitedefaultmidpunct}
{\mcitedefaultendpunct}{\mcitedefaultseppunct}\relax
\EndOfBibitem
\bibitem[Kashizadeh \emph{et~al.}(2015)Kashizadeh, Hambleton, and
  Stanier]{kashizadeh2015numerical}
E.~Kashizadeh, J.~Hambleton and S.~Stanier, Proc. 14th Int. Conf. IACMAG,
  Kyoto, Japan, 2015, pp. 159--164\relax
\mciteBstWouldAddEndPuncttrue
\mciteSetBstMidEndSepPunct{\mcitedefaultmidpunct}
{\mcitedefaultendpunct}{\mcitedefaultseppunct}\relax
\EndOfBibitem
\bibitem[Jin \emph{et~al.}(2020)Jin, Shi, and Hambleton]{jin2020small}
Z.~Jin, Z.~Shi and J.~Hambleton, \emph{Spree Internal Report}, 2020,
  20--7/495S\relax
\mciteBstWouldAddEndPuncttrue
\mciteSetBstMidEndSepPunct{\mcitedefaultmidpunct}
{\mcitedefaultendpunct}{\mcitedefaultseppunct}\relax
\EndOfBibitem
\bibitem[Feng \emph{et~al.}(2019)Feng, Blumenfeld, and Liu]{feng2019support}
Y.~Feng, R.~Blumenfeld and C.~Liu, \emph{Soft matter}, 2019, \textbf{15},
  3008--3017\relax
\mciteBstWouldAddEndPuncttrue
\mciteSetBstMidEndSepPunct{\mcitedefaultmidpunct}
{\mcitedefaultendpunct}{\mcitedefaultseppunct}\relax
\EndOfBibitem
\bibitem[Yue \emph{et~al.}(2018)Yue, Smith, Chen, Chantharayukhonthorn, Kamrin,
  and Grinspun]{yue2018hybrid}
Y.~Yue, B.~Smith, P.~Y. Chen, M.~Chantharayukhonthorn, K.~Kamrin and
  E.~Grinspun, \emph{ACM Transactions on Graphics (TOG)}, 2018, \textbf{37},
  1--19\relax
\mciteBstWouldAddEndPuncttrue
\mciteSetBstMidEndSepPunct{\mcitedefaultmidpunct}
{\mcitedefaultendpunct}{\mcitedefaultseppunct}\relax
\EndOfBibitem
\bibitem[Chen \emph{et~al.}(2021)Chen, Chantharayukhonthorn, Yue, Grinspun, and
  Kamrin]{chen2021hybrid}
P.~Y. Chen, M.~Chantharayukhonthorn, Y.~Yue, E.~Grinspun and K.~Kamrin,
  \emph{Journal of the Mechanics and Physics of Solids}, 2021,  104404\relax
\mciteBstWouldAddEndPuncttrue
\mciteSetBstMidEndSepPunct{\mcitedefaultmidpunct}
{\mcitedefaultendpunct}{\mcitedefaultseppunct}\relax
\EndOfBibitem
\bibitem[Zhu \emph{et~al.}(2017)Zhu, Zhao, Li, Tang, and
  Wang]{zhu2017dynamically}
F.~Zhu, J.~Zhao, S.~Li, Y.~Tang and G.~Wang, Computer Graphics Forum, 2017, pp.
  381--392\relax
\mciteBstWouldAddEndPuncttrue
\mciteSetBstMidEndSepPunct{\mcitedefaultmidpunct}
{\mcitedefaultendpunct}{\mcitedefaultseppunct}\relax
\EndOfBibitem
\bibitem[Steffen \emph{et~al.}(2008)Steffen, Kirby, and
  Berzins]{steffen2008analysis}
M.~Steffen, R.~M. Kirby and M.~Berzins, \emph{International journal for
  numerical methods in engineering}, 2008, \textbf{76}, 922--948\relax
\mciteBstWouldAddEndPuncttrue
\mciteSetBstMidEndSepPunct{\mcitedefaultmidpunct}
{\mcitedefaultendpunct}{\mcitedefaultseppunct}\relax
\EndOfBibitem
\bibitem[Minetti \emph{et~al.}(2009)Minetti, Machtsiras, and
  Masters]{minetti2009optimum}
A.~E. Minetti, G.~Machtsiras and J.~C. Masters, \emph{Journal of biomechanics},
  2009, \textbf{42}, 2188--2190\relax
\mciteBstWouldAddEndPuncttrue
\mciteSetBstMidEndSepPunct{\mcitedefaultmidpunct}
{\mcitedefaultendpunct}{\mcitedefaultseppunct}\relax
\EndOfBibitem
\bibitem[Sidelnik and Young(2006)]{sidelnik2006optimising}
N.~Sidelnik and B.~Young, \emph{Sports Engineering}, 2006, \textbf{9},
  129--135\relax
\mciteBstWouldAddEndPuncttrue
\mciteSetBstMidEndSepPunct{\mcitedefaultmidpunct}
{\mcitedefaultendpunct}{\mcitedefaultseppunct}\relax
\EndOfBibitem
\bibitem[Askari and Kamrin(2016)]{askari2016intrusion}
H.~Askari and K.~Kamrin, \emph{Nature materials}, 2016, \textbf{15},
  1274--1279\relax
\mciteBstWouldAddEndPuncttrue
\mciteSetBstMidEndSepPunct{\mcitedefaultmidpunct}
{\mcitedefaultendpunct}{\mcitedefaultseppunct}\relax
\EndOfBibitem
\end{mcitethebibliography}
\bibliographystyle{rsc} 

\newpage
\onecolumn

\end{document}